\pdfoutput=1

\documentclass[11pt]{article}

\usepackage[preprint]{acl}

\usepackage{times}
\usepackage{latexsym}

\usepackage[T1]{fontenc}

\usepackage[utf8]{inputenc}

\usepackage{microtype}

\usepackage{inconsolata}

\usepackage[para]{threeparttable}
\usepackage{booktabs}
\usepackage{multirow}
\usepackage{graphicx}
\usepackage{amsmath}
\usepackage{amssymb}
\usepackage{hyperref}
\usepackage{mathtools}
\usepackage[ruled,vlined]{algorithm2e}

\newcommand\blfootnote[1]{%
  \begingroup
  \renewcommand\thefootnote{}\footnote{#1}%
  \addtocounter{footnote}{-1}%
  \endgroup
}

\title{Ask Optimal Questions: Aligning Large Language Models with \\ Retriever's Preference in Conversation}

\author{
    Chanwoong Yoon$^{1*}$ \quad Gangwoo Kim$^{1*}$ \quad Byeongguk Jeon$^{2}$ \quad Sungdong Kim$^{2}$ \\ 
    \textbf{Yohan Jo}$^{3}$ \quad \textbf{Jaewoo Kang}$^{1,4\dagger}$ \\
    Korea University$^{1}$ KAIST AI$^{2}$  Seoul National University$^{3}$ AIGEN Sciences$^{4}$ \\
    \texttt{\{cwyoon99, gangwoo\_kim, kangj\}@korea.ac.kr} \\
    \{\texttt{sungdong.kim, byeongguk\}@kaist.ac.kr} \;  \texttt{yohan.jo@snu.ac.kr}
}

\begin{document}
\maketitle

\blfootnote{\textsuperscript{$\ast$} Equal contribution \textsuperscript{$\dagger$} Corresponding author}

\begin{abstract}
Conversational search, unlike single-turn retrieval tasks, requires understanding the current question within a dialogue context.
The common approach of \textit{rewrite-then-retrieve} aims to decontextualize questions to be self-sufficient for off-the-shelf retrievers, but most existing methods produce sub-optimal query rewrites due to the limited ability to incorporate signals from the retrieval results.
To overcome this limitation, we present a novel framework \textsc{RetPO} (\textbf{Ret}riever's \textbf{P}reference \textbf{O}ptimization), which is designed to optimize a language model (LM) for reformulating search queries in line with the preferences of the target retrieval systems.
The process begins by prompting a large LM to produce various potential rewrites and then collects retrieval performance for these rewrites as the retrievers' preferences. 
Through the process, we construct a large-scale dataset called \textsc{RF collection}, containing \textbf{R}etrievers' \textbf{F}eedback on over 410K query rewrites across 12K conversations.
Furthermore, we fine-tune a smaller LM on this dataset to align it with the retrievers' feedback.
Our resulting model demonstrates superiority on two benchmarks, surpassing the previous state-of-the-art performance of \textit{rewrite-then-retrieve} approaches. \footnote{The code and dataset are available at \href{https://github.com/dmis-lab/RetPO}{github.com/dmis-lab/RetPO}}

\end{abstract}

\section{Introduction}

\begin{figure} [t!]
    \centering
    \includegraphics[width=\columnwidth]{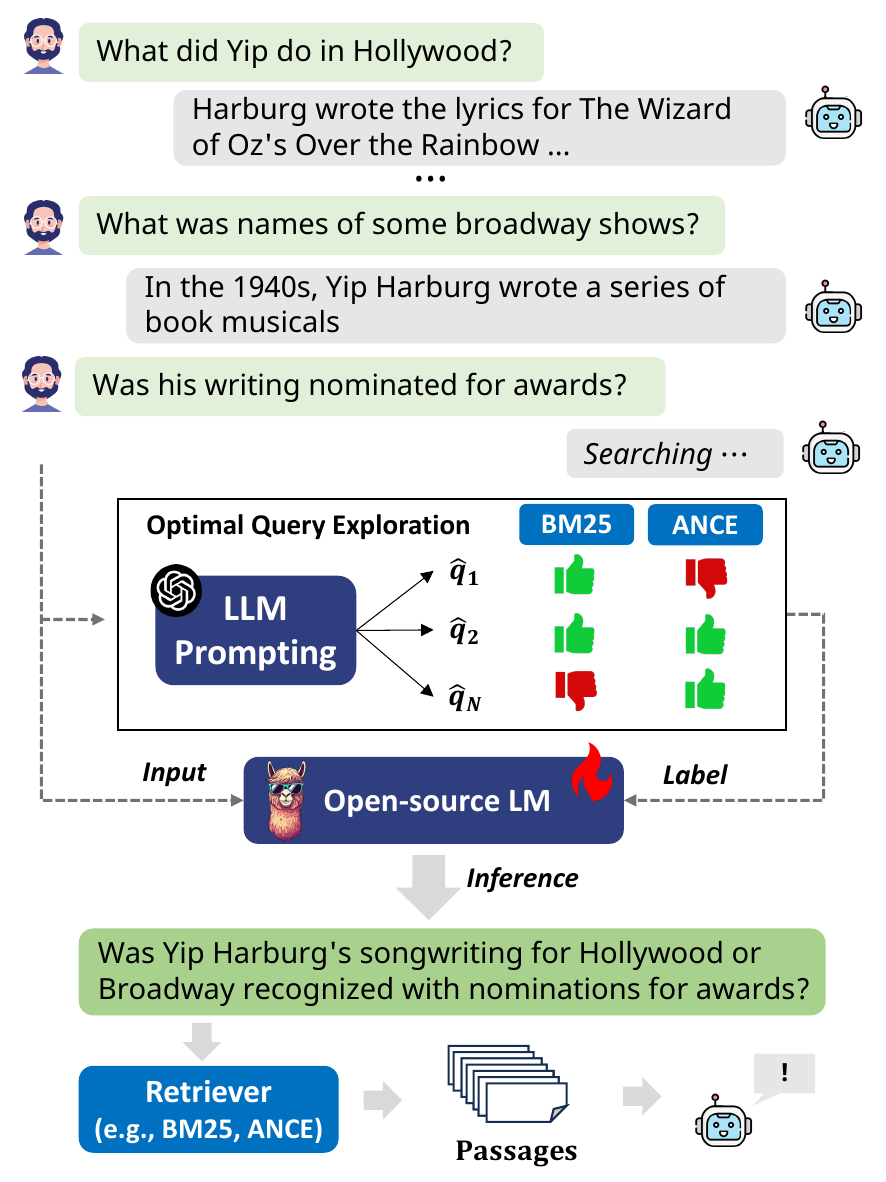}
    \caption{Overview of \textsc{RetPO}. Given a conversation and a follow-up question, (1) potential rewrites $\hat{q}_i$ are generated by prompting an LLM. (2) Retriever's preferences for each rewrite are collected. (3) A smaller LM is trained to be aligned with the retriever's preferences. The resulting model can generate clear and specific rewrites.
    }\vspace{-0.7cm}
    \label{fig:example}
\end{figure}

Conversational search extends the information retrieval to encompass nuances of dialogue context. 
Unlike standard retrieval tasks in open-domain question answering (QA)~\cite{joshi2017triviaqa, kwiatkowski2019natural}, the task is characterized by conversational dependencies in questions (e.g., omission, ambiguity, and coreference)~\cite{qu2020open, anantha2021open, adlakha2022topiocqa}.
As depicted in Figure~\ref{fig:example}, the question in the last turn ``\textit{Was his writing nominated for awards}?'' could only be understood within the context.
Hence, conventional retrieval systems that are not designed to consider dialogue context tend to yield poor retrieval performance.

A prevalent approach to overcome this challenge is \textit{rewrite-then-retrieve}, where questions are decontextualized and made self-contained before being used for retrieval systems.
In many prior works, language models (LMs) are trained for question rewriting (QR) using human rewrites as ground truth~\cite{elgohary2019can,anantha2021open, vakulenko2021question, qian-dou-2022-explicit}.
However, this approach often results in less effective rewrites for search purposes, as human rewrites are typically created without considering their impact on retrieval performance.
Although recent studies~\cite{wu2022conqrr, mo2023convgqr} suggest incorporating signals from retrieval results into the training of QR models, there is still a challenge in fully utilizing the retrievers' preferences across various potential rewrites.

To align a QR model with retrievers' preferences, we present \textsc{RetPO} (\textbf{Ret}riever's \textbf{P}reference \textbf{O}ptimization).
This novel framework aims to optimize a language model (LM) to produce query rewrites tailored to a target retriever's feedback.
\textsc{RetPO} involves several key steps:
(1) we begin with instructing a superior large LM (LLM), GPT-4~\cite{achiam2023gpt}, to provide a variety of potential rewrites with several prompting methods.
(2) We then gather the retriever's feedback on each rewrite (i.e., retrieval performance), 
resulting in a large-scale dataset  \textsc{RF collection}, 
containing \textbf{R}etrievers' \textbf{F}eedback on over 410K query rewrites refined for search purpose across 12K conversations.
(3) Based on our dataset, we further align an open-source LM, such as Llama2-7b~\cite{touvron2023llama}, with preference-driven optimization.
The LM is optimized to generate preferred rewrites over less preferred ones and then is used for the inference phase.

Our experimental results demonstrate that \textsc{RetPO} largely advances retrieval performances on two recent conversational search benchmarks, QReCC~\cite{anantha2021open} and TopiOCQA~\cite{adlakha2022topiocqa}. 
Notably, our 7-billion-parameter model outperforms existing QR approaches, including its teacher model GPT-4.
It also surpasses the previous state-of-the-art performance of BM25 by significant margins 11.8 (MRR) and 19.0 (Recall@10) on QReCC.
Furthermore, we thoroughly analyze our rewrites from \textsc{RF Collection} and \textsc{RetPO}.
The results demonstrate our methods tend to produce specific and detailed rewrites as exemplified in Figure \ref{fig:example}, contributing to the superior retrieval performance.
In GPT-4 evaluation, our rewrites are more favored than human rewrites in terms of clarity and informativeness.

Our contributions are summarized in threefold:
\begin{itemize}
    \item We define optimal query in conversational search and propose how to explore and exploit it. 
    To our knowledge, \textsc{RetPO} is the first to leverage retriever preference-driven optimization for query reformulation. 
    \item  We construct and release \textsc{RF Collection}, a large-scale dataset of \textbf{R}etriever's \textbf{F}eedback on query rewrites in dialogue.
    Our rewrites are superior to human rewrites in retrieval tasks and GPT-4 evaluation.
    \item We align an open-source LM with our dataset. 
    It achieves new state-of-the-art performance of \textit{rewrite-then-retrieve} approaches on two benchmarks, QReCC and TopiOCQA.
\end{itemize}

\begin{figure*} [t]
    \centering
    \includegraphics[width=\textwidth]{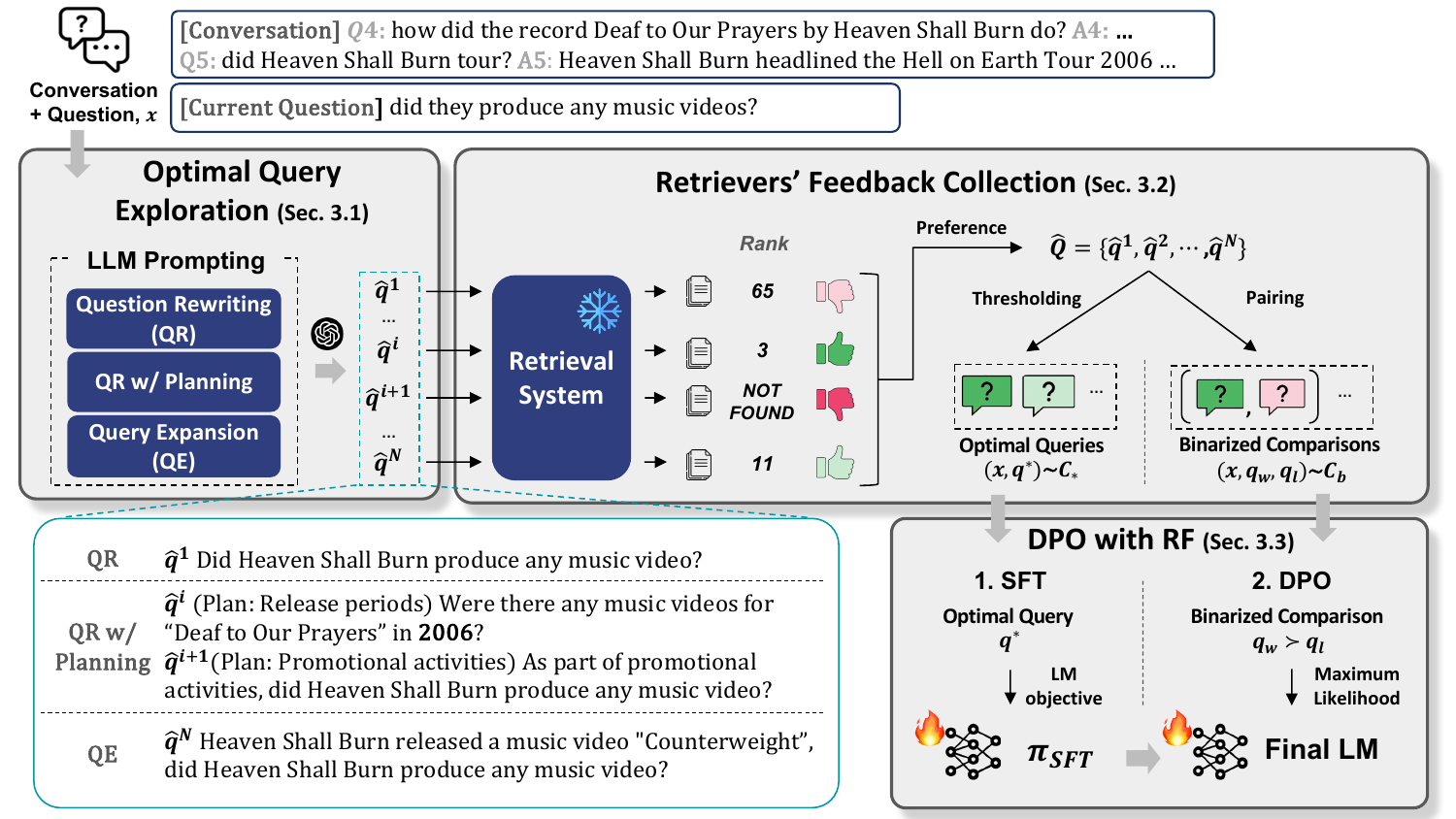}
    \caption{Components of \textsc{RetPO} designed to align an LM with retrievers' preferences. 
    Given a conversation and a user question, we first prompt a capable LLM to provide potential rewrites using various prompting methods (\textit{Optimal Query Exploration}; Sec.~\ref{subsec:oqe}). 
    We then collect the retrievers' feedback on each rewrite by measuring their retrieval performance, leading to two datasets: optimal queries $\mathcal{C}_*$ and query pairs $\mathcal{C}_b$ (\textsc{RF Collection}; Sec.~\ref{subsec:rf_collection})
    Lastly, we optimize an open-source LM with our datasets, encouraging it to generate preferable rewrites (Sec.~\ref{subsec:dpo}). 
    }\vspace{+3mm}
    \label{fig:overview}
\end{figure*}
\section{Background}
\subsection{Task Formulation}
In conversational search, given the current question $q_t$ and the conversation history of question-answer pairs $H_{<t} = \{q_{i}, a_{i} \}^{t-1}_{i=1}$,\footnote{We drop the subscript in the later sections to avoid clutter.} a retrieval system $\text{Ret}(q)$ returns the top-$k$ relevant passages $D_k = \{d_i\}^{k}_{i=1}$ from the target corpus.
In recent \textit{rewrite-then-retrieve} approaches~\cite{anantha2021open, adlakha2022topiocqa},
a question rewriting model $\pi_{QR}$ is trained to generate a self-contained question $q$ by encoding a concatenation of the utterances so far $x=\text{Concat}(H_{<t}, q_t)$; then it predicts a rewrite $\hat{q}$ for use with off-the-shelf retrievers.
Since self-contained rewrites are not always available in natural conversation, most studies rely on the human rewrites for supervision~\cite{elgohary2019can}.

\subsection{Definition of Optimal Query}
Given an evaluation metric $\text{Eval}(\cdot, d^+)$ assessing the retrieved passages based on the gold passage $d^+$, we define optimal query $q^*$ as a query that maximizes the evaluation score as follows:
$$
q^* = \arg \max_{q}~ \text{Eval}(\text{Ret}(q), d^+)
$$
Note that we assume $\text{Ret}(\cdot)$ as frozen.
Under the definition, we argue that previous works using human rewrites as ground truth would result in sub-optimal queries.
The human rewrites are crafted without considering the subsequent retrieval process and its end performance, simply focusing on resolving conversational dependencies.
Although a few studies~\cite{wu2022conqrr, mo2023convgqr} try to incorporate the training signals from the retrieval step, 
they could not exploit training signals from contrasting multiple queries explored with various reasoning types.
\section{Retriever's Preference Optimization}
We newly introduce \textsc{RetPO} (\textbf{Ret}riever’s \textbf{P}reference \textbf{O}ptimization) designed to optimize a query reformulation model with the preference of the retrieval system as illustrated in Figure \ref{fig:overview}.
We first explore a range of potential rewrites with various prompting methods \textit{(Optimal Query Exploration}; Sec. \ref{subsec:oqe}).
We then collect the retriever's feedback on each rewrite by measuring the retrieval performance, resulting in a large-scale dataset, \textsc{RF Collection} (Sec. \ref{subsec:rf_collection}).
By using the dataset, we further align an open-source LM with the preference-driven optimization (Sec. \ref{subsec:dpo}).

\subsection{Optimal Query Exploration}
\label{subsec:oqe}

To explore a broad range of effective search queries, we first prompt a superior LLM to provide a number of potential rewrites.
Based on the conversation and the current question, 
we prompt GPT-4~\cite{achiam2023gpt} with various prompting methods based on different reasoning abilities and purposes.

We adopt three prompting methods: 
(a) \textbf{Question Rewriting}\footnote{See Table \ref{table:prompt_qr} for the question rewriting prompt.} requests the LLM to contextualize the question by resolving coreferences and ellipses.
For example, in Figure \ref{fig:overview}, it finds what a pronoun ``\textit{they}'' in the current question indicates and then replaces it with the exact entity ``\textit{Heaven Shall Burn}'' in the rewrite $\hat{q}^1$.
We initiate our task instruction following \citet{ye2023enhancing} to enhance informativeness and consistency of the rewrite by mentioning `\textit{The resulting question should retain its original meaning and be as informative as possible}.'

Moving beyond resolving the explicit dependencies, 
we devise (b) \textbf{QR with Planning}\footnote{See Table \ref{table:prompt_plan} for the planning prompt.} that allows the LLM to identify an important point to be asked and specify the question's aim.
For example, in Figure \ref{fig:overview}, the rewrite $\hat{q}^i$ inquires about the specific music video and release period mentioned in the conversation.
To this end, it performs an intermediate reasoning step before generating the rewrite, inspired by Chain-of-Thought prompting~\cite{wei2022chain}.
In particular, we encourage the LLM to elicit relevant information from its parametric knowledge or the held-out conversation.

In addition, we adopt (c) \textbf{Query Expansion},\footnote{See Table \ref{table:prompt_qe} for the query expansion prompt.} recently known to be effective in retrieval tasks~\cite{mao2021generation, wang2023query2doc, mo2023convgqr}.
We first instruct the LLM to provide a plausible answer or relevant information without access to external knowledge.
We then append the pseudo-answer to a self-contained rewrite, either a human rewrite if available or the result of the QR prompting method.
As exemplified in Figure \ref{fig:overview}, the rewrite $\hat{q}^N$ is composed of multiple sentences containing the potential answer ``\textit{Counterweight}''.
It increases the chance of keyword overlap between the query and the gold passage, providing informative clues to the retrieval system.
 
With each prompting method, the LLM generates a long text containing from five to ten queries separated by the special token in a single call.
By doing so, it prevents the LLM from generating duplicated queries, resulting in more diverse queries.
As a result, our synthetic queries vary in terms of format and intent.

\subsection{Retrievers' Feedback Collection}
\label{subsec:rf_collection}
Upon the queries collected through the \textit{Optimal Query Exploration}, we gather feedback from target retrievers.
In particular, we feed each query candidate to the frozen retriever and evaluate the outcome.
The retrieval performance is considered as a measurement of the preference.
We use the relative rank of the gold passage in the retrieved passage set.
We eventually construct a synthetic dataset, \textsc{RF Collection}, \textbf{R}etrievers' \textbf{F}eedback on 410K query rewrites across 12K conversations.\footnote{We thoroughly analyze the dataset in Sec. \ref{appendix:eval_dataset}.}

Our dataset consists of two sets, one for supervised fine-tuning and one for preference optimization (discussed in the later section). 
We first construct a collection of optimal queries $\mathcal{C}_*$ under our definition.
Specifically, we choose the five highest-ranked rewrites whose ranks are within a pre-defined threshold.
If all generated queries fail to surpass the threshold, we select the highest-ranked rewrite.
It is used for fine-tuning our model with the language modeling objective, potentially replacing human rewrites.  

For the preference optimization, we construct a collection of binarized comparisons $\mathcal{C}_b$ based on the retriever's feedback.
Given all rewrite candidates for the same input $x$, we first sort them by their rank in ascending order, resulting in $\hat{Q} = \{\hat{q}^1, \hat{q}^2, \cdots, \hat{q}^{|\hat{Q}|}\}$, where the preference becomes $\hat{q}^1 \succ \hat{q}^2 \succ \cdots \succ \hat{q}^{|\hat{Q}|}$.
We then obtain valid pairs of distinct queries $\{(\hat{q}^j, \hat{q}^k): j < k\}$ without duplication of query or rank. 
We randomly sample comparison pairs $(q_w, q_l)$
of `preferred' query $q_w$ and `dispreferred' query $q_l$.
We filter out cases where the preferred query fails to surpass a rank threshold.

\begin{table*}[t!]
    \centering
    \small
    \begin{threeparttable}
    \begin{tabular*}{0.93\textwidth}{cl|cccc|cccc}
        \toprule
        & &   \multicolumn{4}{c}{\textbf{TopiOCQA}} &  \multicolumn{4}{c}{\textbf{QReCC}}  \\ 
        \textbf{Type} & \textbf{Query Reform.} & \textbf{MRR} & \textbf{NDCG} & \textbf{R@10}  & \textbf{R@100} & \textbf{MRR} & \textbf{NDCG} & \textbf{R@10}  & \textbf{R@100} \\
        \midrule \midrule
        \multirow{9}{*}{\rotatebox[origin=c]{90}{\textbf{Sparse (BM25)}}} & \texttt{Original} & 2.1 & 1.8 & 4.0 & 9.1 & 6.5 & 5.6 & 11.1 & 21.5 \\
        & \texttt{Human Rewrite}  & - & - & - & - & 39.8 & 36.3 & 62.7 & 98.5 \\
        \cmidrule(lr){2-10}
        
        & T5QR  & 11.3 & 9.8 & 22.1 & 44.7 & 33.4 & 30.2 & 53.8 & 86.1 \\
        & CONQRR  & - & - & - & - & 38.3 & - & 60.1 & 88.9 \\
        & ConvGQR & 12.4 & 10.7 & 23.8 & 45.6 & 44.1 & 41.0 & 64.4 & 88.0 \\
        & EDIRCS & - & - & - & - & 41.2 & - & 62.7 & \textbf{90.2} \\
        & LLM IQR & - & - & - & - & 49.4 & - & 67.1 & 88.2 \\
        & IterCQR & 16.5 & 14.9 & 29.3 & 54.1 & 46.7 & 44.1 & 64.4 & 85.5 \\
        
        & \textsc{RetPO} (\textit{Ours}) & \textbf{28.3} & \textbf{26.5} & \textbf{48.3} & \textbf{73.1} & \textbf{50.0} & \textbf{47.3} & \textbf{69.5} & 89.5 \\
        \midrule \midrule
        \multirow{9}{*}{\rotatebox[origin=c]{90}{\textbf{Dense (ANCE)}}} & \texttt{Original} & 3.0 & 2.7 & 6.0 & 10.2 & 10.8 & 9.8 & 16.8 & 23.9 \\
        & \texttt{Human Rewrite} & - & - & - & -  & 41.3 & 38.3 & 63.3 & 81.7\\
        \cmidrule(lr){2-10} 
        & T5QR  & 23.0 & 22.2 & 37.6 & 54.4 & 34.5 & 31.8 & 53.1 & 72.8 \\
        & CONQRR  & - & - & - & - & 41.8 & - & 65.1 & 84.7 \\
        & ConvGQR & 25.6 & 24.3 & 41.8 & 58.8 & 42.0 & 39.1 & 63.5 & 81.8 \\
        & EDIRCS & - & - & - & - & 42.1 & - & 65.6 & \textbf{85.3} \\
        & IterCQR & 26.3 & 25.1 & 42.6 & 62.0 & 42.9 & 40.2 & 65.5 & 84.1 \\
        & LLM4CS (\textit{Single}) & 22.6 & 21.2 & 40.1 & 57.5 & 33.2 & 30.8 & 50.3 & 66.0 \\
        & LLM4CS (\textit{Ensemble}) & 32.0 & 31.1 & 50.5 & 67.7 & 41.8 & 39.1 & 62.3 & 76.2  \\
        &  \textsc{RetPO (\textit{Ours})} & \textbf{32.2} & \textbf{31.1} & \textbf{51.6} & \textbf{69.5} & \textbf{44.0} & \textbf{41.1} & \textbf{66.7} & 84.6  \\ 
        \midrule 
        \bottomrule 
    \end{tabular*}
    \end{threeparttable}
    \caption{
    Evaluation results of various retrieval system types on the development sets of QReCC~\cite{anantha2021open} and TopiOCQA~\cite{adlakha2022topiocqa}. 
    We include baselines that integrate the retrievers without fine-tuning.
    In LLM4CS, \textit{Single} and \textit{Ensemble} refer to REW and RAR prompting, respectively.
    }
    \vspace{-2mm}
    \label{table:main}
\end{table*}

\subsection{Direct Preference Optimization with Retrievers' Feedback}
\label{subsec:dpo}

Based on \textsc{RF Collection}, we align a smaller open-source LM with the retriever's preference.
We first fine-tune an LM on the collection of optimal queries in a supervised manner (SFT).
We further align the fine-tuned model with direct preference optimization (DPO)~\cite{rafailov2023direct}.

\paragraph{Supervised Fine-Tuning}
To build an LM that effectively reformulates a question, we fine-tune it in two steps.
The LM is first trained to replicate the ground-truth response following the utterances.
It also aims to benefit the capability to generate pseudo-answers in the query expansion.
We subsequently fine-tune the LM on the optimal queries we collect.
To this end, it learns to generate self-contained and preferable rewrites.
Specifically, we optimize the LM to maximize the log-likelihood for returning the tokens of optimal rewrites $q^{*}$ from the collection $\mathcal{C_*}$.
Given the input $x=\text{Concat}(H_{<t}, q_t)$, the LM $\pi$ is trained as:
$$
    \pi_{SFT} = \max_{\pi} \mathbb{E}_{(x, q^*) \sim \mathcal{C_*}} \log \pi(q^* \mid x)
$$

\paragraph{Direct Preference Optimization}
Initiating with the SFT model, we further align the LM with the retrievers' preferences.
In particular, we apply DPO, a method recently highlighted by \citet{rafailov2023direct}, for its efficacy in alignment learning. 
It optimizes the student model $\pi_{\theta}$ to maximize the likelihood of generating the preferred $q^{w}$ over $q^{l}$, starting from the $\pi_{SFT}$.
$$
    J(\theta)=\mathbb{E}_{(x, q_w, q_l) \sim \mathcal{C}_b} \log \sigma (r_{\theta}(x, q_w) - r_{\theta}(x, q_l))
$$

Following \citet{rafailov2023direct}, we simplify $r_{\theta}(x, q) = \beta \log \pi(q \mid x) - \beta \log \pi_{SFT}(q \mid x)$ with the likelihood difference with the
SFT model.

This process is guided by the principle of maximizing the contrast between preferred and dispreferred rewrites, thereby providing a clear signal for model training.
DPO enables the model to directly learn from the contrast by focusing on the relative merits of each rewrite as judged by the retrieval system. 
Through this targeted optimization, the SFT model is further trained to generate rewrites that reflect the nuanced preferences of the target retriever.




\section{Experiment}

\paragraph{Datasets}
We test our models on two recent open-domain CQA benchmarks, QReCC~\cite{anantha2021open} and TopiOCQA~\cite{adlakha2022topiocqa}.
QReCC contains 14K conversations with 81K question-answer pairs and self-contained rewrites.
TopiOCQA is a more recent benchmark consisting of 3.9K conversations, presenting a challenge of topic switches.
To test the models in the \textit{zero-shot} setup, we also include CAsT-20~\cite{dalton2021cast} that does not contain the train set\footnote{See more details in Appendix \ref{append:dataset}}.

\paragraph{Retrieval Systems}
To investigate the impact of different types of retrieval systems, we adopt a sparse retriever BM25 and dense retrievers ANCE~\cite{xiong2020approximate} and Contriever~\cite{izacard-grave-2021-leveraging}, widely used in the task. 
Specifically, we use the checkpoints trained on MS-MARCO~\cite{bajaj2016ms} passage retrieval task. 
Note that we do not further fine-tune the retrievers for our target task.

\paragraph{Evaluation Metrics}
We use several evaluation metrics, following previous works. 
Mean Reciprocal Rank (\textbf{MRR}) is the average of the ranks measuring how effectively the retriever can locate gold passages. 
Normalized Discounted Cumulative Gain (\textbf{NDCG@3}) evaluates retrieval results by considering both relevance and rank of top-3 results. 
\textbf{Recall@$k$} verifies whether the retriever succeeds in locating gold passages within top-$k$ results. 

\begin{table}[t!]
    \small
    \centering
    \begin{threeparttable}
    \begin{tabular*}{1\columnwidth}{lccc}
        \toprule
        & \multicolumn{3}{c}{\textbf{TopiOCQA}} \\
        \textbf{Query Reformulation} &  \textbf{MRR} & \textbf{R@10}  & \textbf{R@100} \\
        \midrule \midrule
        GPT-4 Prompting (Teacher)  & 18.5 & 35.1 & 62.9 \\
        Distillation to Llama2-7b & 19.0 & 35.5 & 64.6 \\
        \midrule
        \textsc{RetPO} (\textit{Ours}) & 28.3 & 48.3 & 73.1 \\
        \quad w/o. DPO & 23.4 & 41.6 & 67.7 \\
        \quad w/o. Query Expansion & 22.0 & 40.2 & 68.5 \\
        \quad w/o. QE and Planning & 21.8 & 39.2 & 67.7 \\
        \midrule\
    \end{tabular*}
    \end{threeparttable}
    \caption{
    Ablation study for each component of \textsc{RetPO}.
    We compare the baselines that prompt GPT-4 to generate the rewrites and fine-tune smaller LM on them.
    }
    \label{table:ablation}
    \vspace{-6mm}
\end{table}
\begin{table*}[t!]
    \centering
    \small
    \begin{threeparttable}
    \begin{tabular*}{0.96\textwidth}{l|ccc|ccc|ccc}
        \toprule
        &  \multicolumn{9}{c}{\textbf{TopiOCQA}}  \\ 
        &  \multicolumn{3}{c}{\textbf{First}} &  \multicolumn{3}{c}{\textbf{Topic-concentrated}} & \multicolumn{3}{c}{\textbf{Topic-shifted}}  \\ 
        \textbf{Query Reformulation} & \textbf{MRR} & \textbf{R@10}  & \textbf{R@100} & \textbf{MRR} & \textbf{R@10}  & \textbf{R@100} & \textbf{MRR} & \textbf{R@10}  & \textbf{R@100} \\
        \midrule \midrule
        \texttt{Original}  & 14.7 & 29.3 & 64.4 & 0.9 & 1.7 & 4.2 & 1.1 & 1.9 & 4.2 \\
        Fine-tuned T5  & 14.7 & 29.3 & 64.4 & 14.4 & 28.2 & 52.4 & 9.4 & 18.2 & 36.9 \\
        GPT-4 Prompting & 15.6 & 31.2 & 62.0 & 19.7 & 37.2 & 65.3 & 16.4 & 31.3 & 57.4 \\
        Distillation to Llama2-7b & 17.9 & 34.2 & 63.9 & 20.0 & 37.1 & 66.3 & 17.0 & 32.0 & 60.7 \\

         \midrule
        \textsc{RetPO} (\textit{Ours}) & \textbf{32.0} & \textbf{51.7} & \textbf{75.1} & \textbf{27.4} & \textbf{47.1} & \textbf{72.4} & \textbf{29.6} & \textbf{50.0} & \textbf{74.3} \\

        \bottomrule 
    \end{tabular*}
    \end{threeparttable}
    \caption{
    Breakdown evaluation of BM25 on the development set of TopiOCQA, segmented by question type: initial turn (\textbf{First}), topic-consistent turns with their preceding one (\textbf{Topic-Concentrated}) and topic-switched turns (\textbf{Topic-Shifted}).
    Following \citet{adlakha2022topiocqa}, we identify a switch of topic if the gold passage is based on a different Wikipedia document.
    }
    \label{table:topic-shift}
    \vspace{-6mm}
\end{table*}

\paragraph{Baselines}
We select several baselines from \textit{rewrite-then-retrieve} approaches.
(1) T5QR~\cite{lin2020conversational} fine-tunes T5-base~\cite{raffel2020exploring} to replicate human rewrites.
(2) CONQRR~\cite{wu2022conqrr} introduces a reinforcement learning framework that leverages retrieval performance as a reward signal.
(3) ConvGQR~\cite{mo2023convgqr} fine-tunes QR models with an auxiliary loss function for injecting the knowledge of the target retriever.
(4) EDIRCS~\cite{mao2023search} extracts tokens from the dialogue and adds a few newly generated tokens.
(5) IterCQR~\cite{jang2023itercqr} incorporates iterative training of QR model driven by query rewrites explored with GPT-3.5.
(6) LLM IQR~\cite{ye2023enhancing} prompts GPT-3.5 multiple times to reformulate questions according to pre-determined criteria.
LLM4CS~\cite{mao-etal-2023-large} proposes three ensemble methods that generate unified query representations by aggregating rewrites and hypothetical responses obtained from multiple inferences of GPT-3.5. We report two variants: the basic approach, REW prompting with MaxProb aggregation, and the best-performing method, RAR prompting with Mean aggregation (refer to \textit{Single} and \textit{Ensemble}, respectively).

We also include baselines that fine-tune retrievers in Sec.~\ref{subsec:compare_cdr}, such as ConvDR~\cite{yu2021few}, ConvANCE and LeCoRE~\cite{mao2023learning}.
The most recent study, InstructoR~\cite{jin-etal-2023-instructor}, instructs GPT-3.5 to augment the train set for fine-tuning the retriever.
\subsection{Main Results}
Table \ref{table:main} shows the evaluation results of various types of retrieval systems.

\textbf{Leveraging signal from the retriever enhances the end performance.}
Encoding the current question without modification (\texttt{Original}) performs poorly.
Performance of T5QR using the human rewrites as supervision is bounded by its label (\texttt{Human Rewrite}).
Other baselines using the same backbone with signals from retrievers (CONQRR, ConvGQR, and IterCQR) largely advance performances on QReCC but struggle with TopiOCQA, implying that TopiOCQA is more complex and challenging than QReCC.

\textbf{While baselines with GPT-3.5 show competitive performances, our 7-billion-parameter model surpass them.}
Our model outperforms or competes consistently against baselines that utilize the much larger QR model, GPT-3.5 (LLM IQR, LLM4CS).
Moreover, despite LLM4CS relying on multiple inference calls and directly accessing extensive world knowledge of GPT-3.5, our approach achieves higher scores with a single inference from a much smaller LM\footnote{See Appendix~\ref{appendix:llm4cs} for a detailed comparison with LLM4CS.}.
This indicates that our model has effectively learned to generate rewrites that are more effective for and preferable to the retriever.


\textbf{\textsc{RetPO} achieves new state-of-the-art performances in most settings}.
Notably, for TopiOCQA, it advances the previous state-of-the-art of BM25 with a prominent gap; 11.8, 19.0, and 19.0 in MRR, R@10, and R@100, respectively.
In the other benchmark and retriever type, \textsc{RetPO} similarly outperforms the prior best results. 
The only exception is R@100 scores\footnote{We observe \textsc{RetPO} sacrifices R@100 score due to its tendency to produce longer and detailed rewrites. See Appendix \ref{appendix:over_spec} for case study.} on QReCC known to exhibit the shortcut between the held-out conversation and the gold passage~\cite{kim2022saving}.
It could make the token extraction method from the conversation (EDIRCS) perform better.
Overall, \textsc{RetPO} shows a consistent improvement over other models across both sparse and dense retrieval systems. 
These results suggest that \textsc{RetPO} highlights the potential of preference-driven training in tailoring more favorable rewrites in various environments.

\begin{figure} [t!]
    \centering
    \includegraphics[width=\columnwidth]{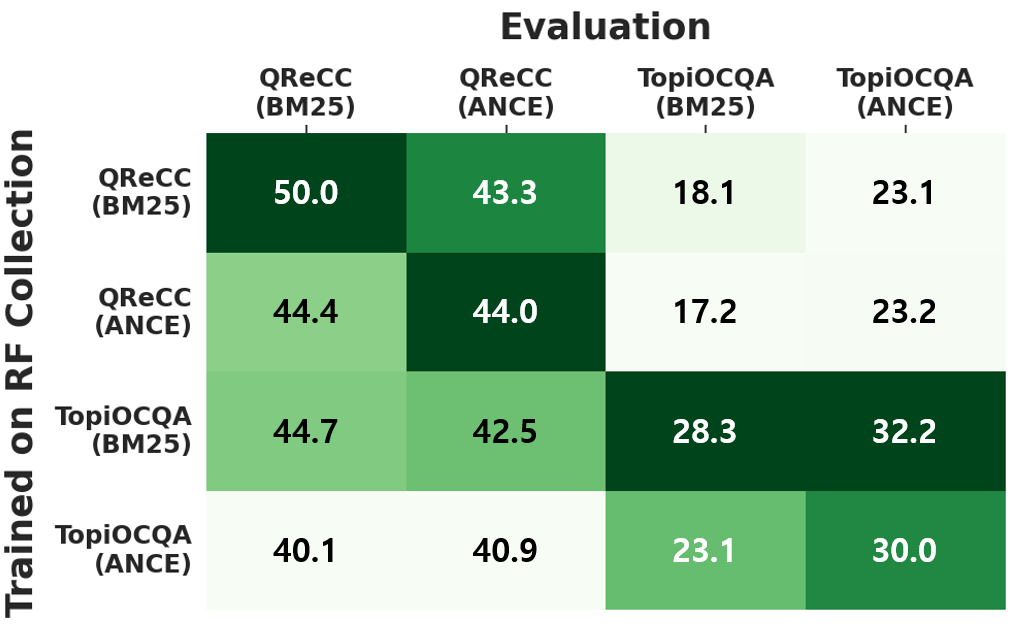}
    \caption{Heatmap of MRR scores when generalizing toward different settings.
    The shades are normalized per column to depict relative performance
    }\vspace{-0.6cm}
    \label{fig:cross_eval}
\end{figure}
\begin{table}[t!]
    \small
    \centering
    \begin{threeparttable}
    \begin{tabular*}{1.00\columnwidth}{lcccc}
        \toprule
        \textbf{Query Reform.} & \textbf{\#Qs} &  \textbf{MRR} & \textbf{R@10}  & \textbf{R@100} \\
        \midrule \midrule
        \multicolumn{5}{c}{\textit{Dense (ANCE)}} \\
        \midrule
        \texttt{Original} & 1 & 10.2 & 15.7 & 22.7 \\
        \texttt{Concat} ($H_{<t}, q_t$) & 1 & 42.8 & 63.7 & 79.9 \\
        \texttt{Human Rewrite} & 1 & 41.3 & 63.3 & 81.7 \\
        \quad \texttt{+ Gold Answer} & 1 & 57.8 & 79.3 & 90.1 \\
        GPT-4 Prompting & 1 & 40.4 & 61.7 & 79.7 \\
        \midrule
        \multicolumn{4}{l}{\textsc{RF Collection}} \\
        \quad Ques. Rewriting & 10 & 57.4 & 75.1 & 87.9 \\
        \quad QR w/ Planning & 10 & 61.7 & 78.9 & 89.8 \\
        \quad Query Expansion & 5 & 62.2 & 81.3 & 92.3 \\
        \quad \texttt{Union} & 25 & 73.6 & 86.8 & 94.5 \\
        \midrule\
    \end{tabular*}
    \end{threeparttable}
    \vspace{-0.5cm}
    \caption{
    Effectiveness of optimal queries in \textsc{RF Collection}. 
    We generate a certain number (\textbf{\#Qs}) of rewrites using each method and report the best retrieval performances among them.
    }
    \label{table:oqf}
    \vspace{-6mm}
\end{table}

\subsection{Ablation Study}
Table \ref{table:ablation} shows ablation results for \textsc{RetPO} on TopiOCQA, by removing its components gradually. 
We start with simple baselines that prompt our teacher model GPT-4 to generate rewrites (row 1), and then fine-tune the smaller LM Llama2-7B on them (row 2).
\textsc{RetPO} (row 3) significantly outperforms the baselines by using preference-driven optimization as useful supervision. 
Without DPO (row 4), the performance drops, indicating the importance of integrating the retriever's preferences for certain rewrites over others. 
Similarly, omitting prompting methods (Query Expansion and Planning) from \textsc{RF Collection} (rows 5 and 6) results in degraded performance, underscoring their contribution to exploring optimal queries. 
The degradation across ablation clearly shows that every component of \textsc{RetPO} is crucial for its superior results in conversational search tasks.
\subsection{Robustness to Topic Shifts in Dialogues}
We report the results segmented by the question types in Table \ref{table:topic-shift}.
We delve into the unique challenge, topic-switching, posed within TopiOCQA, where topics may abruptly change between turns. 
\textsc{RetPO} exhibits exceptional robustness in handling these topic shifts, significantly outperforming baselines. 
Its performances on \textit{Topic-shifted} queries are even higher than those on \textit{Topic-concentrated} queries, in contrast to the tendency of the baselines.
\footnote{This improvement might be related to \textsc{RetPO}'s tendency to specify details . See Appendix \ref{subsec:q_dist} for detailed analysis}
Additionally, \textsc{RetPO} boosts performance even on the context-independent queries (\textit{first}), suggesting its potential for enhancing single-turn retrieval tasks as well.

\subsection{Generalizing to Different Preferences}
In Figure \ref{fig:cross_eval}, we explore how well models generalize across datasets with varying scenarios. 
The performances along the heatmap's diagonal reveal that models typically excel when the dataset and the retriever are the same between training and evaluation, as expected. 
For TopiOCQA, however, we observe that the model aligned with BM25 performs better even when ANCE is used for evaluation. This might be linked to the effectiveness of query expansion strategies more favored by BM25.\footnote{See Appendix \ref{subsec:q_dist} for detailed analysis}
Additionally, models generalize relatively well from TopiOCQA to QReCC, compared to the opposite direction.
It again indicates that the challenges posed by TopiOCQA are more complex than QReCC.
Furthermore, the results showcase the potential utility of our method to identify and select the most effective combination of strategies.
\section{Analysis and Discussion}
\begin{table*}[t!]
    \small
    \centering
    \begin{threeparttable}
    \begin{tabular*}{0.85\textwidth}{lcccccc}
        \toprule
        \textbf{Model} & \textbf{LM} &  \textbf{Retriever} & \textbf{TopiOCQA} &  \textbf{QReCC}  & \textbf{CAsT-19} & \textbf{CAsT-20} \\
        \midrule \midrule
        \multicolumn{7}{c}{\textbf{\textit{With Fine-tuning of Retriever}}} \\
        \midrule
        ConvDR & - & ANCE  & 26.4 & 35.7 & 43.9 & 32.4 \\
        Conv-ANCE & - & ANCE  & 20.5 & 45.6 & 34.1 & 27.5 \\
        HAConvDR & - & ANCE  & 28.5 & 45.6 & - & - \\
        LeCoRE & - & SPLADE  & 31.4 & \underline{48.5} & 42.2 & 29.0 \\
        InstructoR & GPT-3.5 & ANCE  & 23.7 & 40.5 & 46.6 & 29.6 \\
         & GPT-3.5 & Contriever  & \textbf{37.0} & \textbf{50.4} & \textbf{55.1} & 32.8 \\
        \midrule \midrule
        \multicolumn{7}{c}{\textbf{\textit{Without Fine-tuning of Retriever}}} \\
        \midrule
        T5QR & T5-base & ANCE  & 22.2 & 31.8 & 41.7 & 29.9 \\
        ConvGQR & T5-base  & ANCE  & 24.3 & 39.1 & 43.4 & 33.1 \\
        LLM4CS (\textit{Single}) & GPT-3.5  & ANCE  & 21.2 & 30.8 & 43.1 & 35.7 \\
        LLM4CS (\textit{Ensemble}) & GPT-3.5  & ANCE  & 31.1 & 39.1 & \underline{51.5} & \textbf{45.5} \\
        \midrule 
        \multicolumn{7}{l}
        {\textsc{RetPO}$_{\text{BM25}}$ (\textit{Ours})} \\
         & Llama2-7B & ANCE  & 31.1 & 40.4 & 44.8 & 36.9 \\
         & Llama2-7B & Contriever  & \underline{33.7} & 45.7  & 42.8 & \underline{41.1} \\
        \midrule\
    \end{tabular*}
    \end{threeparttable}
    \caption{
    Evaluation results of our method and the baselines that fine-tunes the retrievers in NDCG@3 scores. \textbf{The best scores} are in bold and \underline{the second-best scores} are underlined. We use the rewrites aligned with BM25 feedback from \textsc{RetPO}$_{\text{BM25}}$. 
    }
    \label{table:sota}
    \vspace{-2mm}
\end{table*}

\subsection{GPT-4 Evaluation}
\begin{figure} [t!]
    \centering
    \includegraphics[width=\columnwidth]{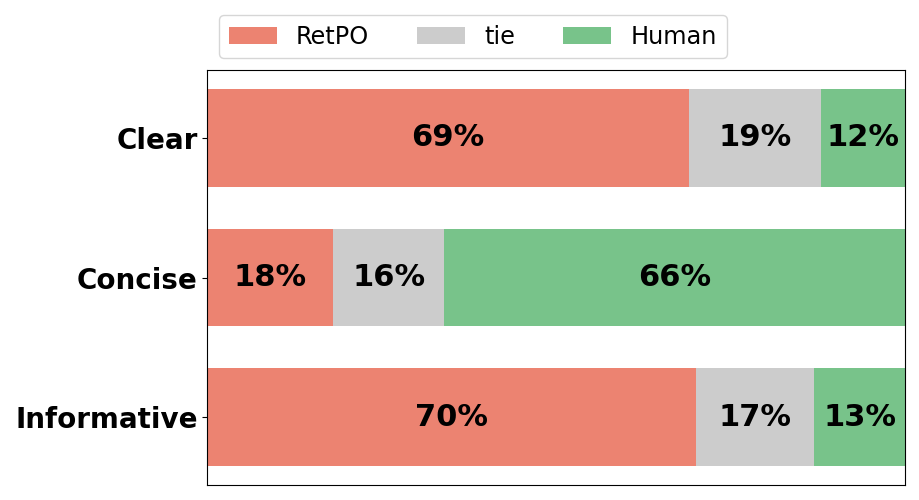}
    \caption{Pairwise evaluation with GPT-4. \textsc{RetPO}'s rewrites are compared with the human rewrites.}\vspace{-0.5cm}
    \label{fig:gpt4_eval_retpo}
\end{figure}
In Figure~\ref{fig:gpt4_eval_retpo}, we perform an automatic pairwise comparison, contrasting queries in three criteria: clarity, conciseness, and informativeness. 
To this end, we randomly sample 100 examples in the validation set and leverage a superior LLM, GPT-4~\cite{achiam2023gpt}, as a judge. 
For the same input conversation, we pair query rewrites from a human and \textsc{RetPO} for comparison.
Figure~\ref{fig:gpt4_eval_retpo} indicates that \textsc{RetPO} typically generates question rewrites that are more informative and less ambiguous compared to human rewrites, though they are less concise. 
The extended rewrites from \textsc{RetPO}, despite sacrificing conciseness, contain valuable details, leading to superior performance. 
We observe a similar tendency for \textsc{RF Collection}.\footnote{More details are in Appendix~\ref{appendix:GPT-4_eval_details} and Appendix~\ref{appendix:eval_dataset}.}

\begin{table}[t!]
    \small
    \centering
    \begin{threeparttable}
    \begin{tabular*}{1\columnwidth}{lccc}
        \toprule
        & \multicolumn{3}{c}{\textbf{QReCC}} \\
        \textbf{Method} &  \textbf{MRR} & \textbf{R@10}  & \textbf{R@100} \\
        \midrule \midrule
        \texttt{Human Rewrite}  & 41.3 & 63.3 & 81.7 \\
        \textsc{RF Collection} & 73.6 & 86.8 & 94.5 \\
        \midrule
        T5QR~\cite{lin2020conversational} & 34.5 & 53.1 & 72.8 \\
        Fine-tuned T5 & & & \\
        \quad w/ \texttt{Human Rewrite} & 35.5 & 55.4 & 73.8 \\
        \quad w/ \textsc{RF Collection} & 39.2 & 59.9 & 78.2 \\
        \midrule\
    \end{tabular*}
    \end{threeparttable}
    \caption{
    Evaluation results of a small language model (T5-base) on the QReCC benchmark. 
    We fine-tune T5-base using two datasets—\texttt{Human Rewrite} and \textsc{RF Collection}—and compare the retrieval performance based on the rewrites generated by each model.
    }
    \label{table:t5-rf}
    \vspace{-6mm}
\end{table}
\subsection{\textsc{RF Collection} with Small LM}
Table~\ref{table:t5-rf} presents experimental results with a small LM, T5-base to assess the effectiveness of \textsc{RF Collection} beyond our proposed method.
To examine its utility as a general training resource, we fine-tune a non-LLM model, T5-base, on \textsc{RF Collection} and evaluate its performance in retrieval tasks by using the dense retriever ANCE.
Notably, T5 models trained on \textsc{RF Collection} achieve substantial improvements over those trained on \texttt{Human Rewrite}, surpassing both the T5QR baseline and our reproduced scores.
These findings underscore \textsc{RF Collection} as a valuable resource for query reformulation, extending its impact beyond large-scale LLM-based methods and reinforcing its role as a standalone contribution to the field.

\subsection{Fine-tuning Retriever or Rewriter?}
\label{subsec:compare_cdr}

Table~\ref{table:sota} compares our method with the baselines that fine-tune retrievers in in-domain (QreCC, TopiOCQA, and CAsT-19) and out-of-domain (CAsT-20) scenarios. 
Fine-tuning retrievers generally yields good in-domain performance but the performance tends to be highly sensitive to retrievers. 
For example, InstructoR is effective for fine-tuning Contriever but its effectiveness drops substantially for a different retriever ANCE, showing the instability of the method.
This highlights the difficulty of fine-tuning retrievers, which requires sophisticated engineering and retriever-specific optimization.

\textbf{Our method consistently enhances the performance of off-the-shelf retrievers without additional feedback for the target retrievers.} 
By utilizing rewrites aligned with BM25, \textsc{RetPO}${_\text{BM25}}$ surpasses existing baselines except for InstructoR on TopiOCQA. 
It particularly shows superior effectiveness in the \textit{zero-shot} scenario.
On the CAsT-20 dataset, which lacks a training set, \textsc{RetPO}${_\text{BM25}}$ trained on TopiOCQA successfully generalizes to the unseen dataset, outperforming all baselines except for LLM4CS which requires multiple LLM inferences\footnote{See Appendix~\ref{appendix:llm4cs} for a detailed comparison with LLM4CS.}.
Our findings implicate that \textsc{RetPO} is easy to deploy and shows competitive performance across diverse scenarios.

\section{Related Works}
\paragraph{Conversational Search}
Conversational search is the precedent task of open-domain conversational QA and several benchmarks are released~\cite{qu2020open, dalton2020cast}.
A line of studies proposes to fine-tune the dense retriever, enabling it to encode conversational context.
Most studies follow the approach~\cite{lin2021contextualized, kim2022generating, mao2022convtrans, ma2023query, huang-etal-2023-converser, mo2024history, mo2024convsdg, chen-etal-2024-generalizing}.
Concurrently, \citet{mao2024chatretriever} proposes to use LLM as a retrieval backbone and achieve superior performance in the task.
Although they show the dominant performances in the task, they require retriever-specific engineering.

\paragraph{Query Reformulation}
Recent studies prompt LMs to provide detailed information such as the expected document~\cite{wang2023query2doc, jagerman2023query}.
The recent study propose to use reward signals to optimize the QR model \cite{ma2023query}. 
In conversational search, \citet{anantha2021open} introduces the \textit{rewrite-then-retrieve} pipeline to handle the conversational dependency.
Most studies fine-tune QR models to generate the standalone question~\cite{voskarides2020quretec, lin2021multistage, kumar2020making}.
Recently, commercial models have been employed for QR tasks, leveraging their robust generative capabilities and extensive world knowledge~\cite{ye2023enhancing, mao-etal-2023-large, mo2024chiq}.
In contrast, \textsc{RetPO} is the first to utilize preference-driven optimization for reformulating queries in conversational search.

\paragraph{Aligning Language Models with Feedback}

Studies on LLM alignment utilize human feedback~\cite{bai2022training, ouyang2022training, rafailov2023direct}. Recently, AI feedback is also actively explored as an alternative to human feedback~\cite{bai2022constitutional, sun2023salmon}.  \citet{kim2023aligning} automatically construct synthetic feedback, leveraging prior knowledge, instead of collecting feedback. 
\citet{tian2023fine} obtain synthetic feedback utilizing truthfulness measurements like FactScore~\cite{min2023factscore}. 
Our method is similar to these studies in that it includes the synthetic dataset construction; however, we focus on a specific target task, question rewriting, and reflecting a target retriever's feedback.
Concurrently, to reduce the reliance on labeled data for aligning QR models, \citet{zhang2024adaptive} utilize marginalized rewards derived from conversation answers, while \citet{lai2024adacqr} leverage the LM as a reference-free evaluator to align retrievers with only a small set of training data.
\section{Conclusion}
Our paper introduces \textsc{RetPO}, a framework for optimizing an LM to generate retriever-preferred query rewrites. 
Utilizing the LLM-based process, we construct and release a large-scale dataset \textsc{RF Collection}.
Based on it, we enhance an open-source LM, significantly outperforming \textit{rewrite-then-retrieve} baselines on two recent benchmarks QReCC and TopiOCQA. 
Our work, which pioneers preference-driven optimization in query reformulation advances conversational search performance and shows promising results in generalization. 
\section*{Limitation}
One limitation of our study is the exclusive focus on larger-scale language models. Consequently, our model tends to generate longer queries rich in specific information and keywords, possibly relying on the emergent abilities of large LMs, which we leverage to boost performance. However, exploring smaller-scale LMs could offer insights into the scalability and efficiency of our approach.

Additionally, due to budget constraints, we utilized only half of the TopiOCQA training set. Access to the full dataset could potentially yield further improvements in model performance.

Our framework has been tested solely within the realm of conversational search, yet its application is not limited to this task. Future research could adapt our framework to a broader range of tasks and domains, potentially enhancing its utility and impact.

While we employed three prompting methods, there is a vast landscape of alternative approaches that we did not explore. Future studies could investigate additional prompting strategies tailored to specific tasks and retriever systems.

Finally, pairing our method with more advanced retrieval systems presents a promising avenue for research. Despite the clarity and consistency of the generated queries, we noted instances of retrieval failure, indicating that there is room for improvement in retriever performance, which could, in turn, further enhance the overall efficacy of our method.
\section*{Acknowledgement}
This work was supported by the National Research Foundation of Korea [NRF2023R1A2C3004176, RS-2023-00262002], the Ministry of Health \& Welfare, Republic of Korea [HR20C0021], and the ICT Creative Consilience program through the Institute of Information \& Communications Technology Planning \& Evaluation (IITP) grant funded by the Korea government (MSIT) [IITP-2025-20200-01819].
This work was also supported by the National Research Foundation of Korea (NRF) grant funded by the Korea government (MSIT) (No. RS-2024-00414981) and Institute of Information \& communications Technology Planning \& Evaluation (IITP) under the Leading Generative AI Human Resources Development (IITP-2025-RS-2024-00397085) grant funded by the Korea government (MSIT).

\bibliography{custom}

\begin{thebibliography}{61}
\expandafter\ifx\csname natexlab\endcsname\relax\def\natexlab#1{#1}\fi

\bibitem[{Adlakha et~al.(2022)Adlakha, Dhuliawala, Suleman, de~Vries, and Reddy}]{adlakha2022topiocqa}
Vaibhav Adlakha, Shehzaad Dhuliawala, Kaheer Suleman, Harm de~Vries, and Siva Reddy. 2022.
\newblock Topiocqa: Open-domain conversational question answering with topic switching.
\newblock \emph{Transactions of the Association for Computational Linguistics}, 10:468--483.

\bibitem[{Anantha et~al.(2021)Anantha, Vakulenko, Tu, Longpre, Pulman, and Chappidi}]{anantha2021open}
Raviteja Anantha, Svitlana Vakulenko, Zhucheng Tu, Shayne Longpre, Stephen Pulman, and Srinivas Chappidi. 2021.
\newblock Open-domain question answering goes conversational via question rewriting.
\newblock In \emph{Proceedings of the 2021 Conference of the North American Chapter of the Association for Computational Linguistics: Human Language Technologies}, pages 520--534.

\bibitem[{Bai et~al.(2022{\natexlab{a}})Bai, Jones, Ndousse, Askell, Chen, DasSarma, Drain, Fort, Ganguli, Henighan et~al.}]{bai2022training}
Yuntao Bai, Andy Jones, Kamal Ndousse, Amanda Askell, Anna Chen, Nova DasSarma, Dawn Drain, Stanislav Fort, Deep Ganguli, Tom Henighan, et~al. 2022{\natexlab{a}}.
\newblock Training a helpful and harmless assistant with reinforcement learning from human feedback.
\newblock \emph{arXiv preprint arXiv:2204.05862}.

\bibitem[{Bai et~al.(2022{\natexlab{b}})Bai, Kadavath, Kundu, Askell, Kernion, Jones, Chen, Goldie, Mirhoseini, McKinnon et~al.}]{bai2022constitutional}
Yuntao Bai, Saurav Kadavath, Sandipan Kundu, Amanda Askell, Jackson Kernion, Andy Jones, Anna Chen, Anna Goldie, Azalia Mirhoseini, Cameron McKinnon, et~al. 2022{\natexlab{b}}.
\newblock Constitutional ai: Harmlessness from ai feedback.
\newblock \emph{arXiv preprint arXiv:2212.08073}.

\bibitem[{Bajaj et~al.(2016)Bajaj, Campos, Craswell, Deng, Gao, Liu, Majumder, McNamara, Mitra, Nguyen et~al.}]{bajaj2016ms}
Payal Bajaj, Daniel Campos, Nick Craswell, Li~Deng, Jianfeng Gao, Xiaodong Liu, Rangan Majumder, Andrew McNamara, Bhaskar Mitra, Tri Nguyen, et~al. 2016.
\newblock Ms marco: A human generated machine reading comprehension dataset.
\newblock \emph{arXiv preprint arXiv:1611.09268}.

\bibitem[{Chen et~al.(2024)Chen, Dou, Mao, Liu, and Zhao}]{chen-etal-2024-generalizing}
Haonan Chen, Zhicheng Dou, Kelong Mao, Jiongnan Liu, and Ziliang Zhao. 2024.
\newblock \href {https://doi.org/10.18653/v1/2024.acl-long.149} {Generalizing conversational dense retrieval via {LLM}-cognition data augmentation}.
\newblock In \emph{Proceedings of the 62nd Annual Meeting of the Association for Computational Linguistics (Volume 1: Long Papers)}, pages 2700--2718, Bangkok, Thailand. Association for Computational Linguistics.

\bibitem[{Dalton et~al.(2021)Dalton, Xiong, and Callan}]{dalton2021cast}
Jeffrey Dalton, Chenyan Xiong, and Jamie Callan. 2021.
\newblock Cast 2020: The conversational assistance track overview.
\newblock In \emph{In Proceedings of TREC}.

\bibitem[{Dalton et~al.(2020)Dalton, Xiong, Kumar, and Callan}]{dalton2020cast}
Jeffrey Dalton, Chenyan Xiong, Vaibhav Kumar, and Jamie Callan. 2020.
\newblock Cast-19: A dataset for conversational information seeking.
\newblock In \emph{Proceedings of the 43rd International ACM SIGIR Conference on Research and Development in Information Retrieval}, pages 1985--1988.

\bibitem[{Elgohary et~al.(2019)Elgohary, Peskov, and Boyd-Graber}]{elgohary2019can}
Ahmed Elgohary, Denis Peskov, and Jordan Boyd-Graber. 2019.
\newblock Can you unpack that? learning to rewrite questions-in-context.
\newblock In \emph{Proceedings of the 2019 Conference on Empirical Methods in Natural Language Processing and the 9th International Joint Conference on Natural Language Processing (EMNLP-IJCNLP)}, pages 5918--5924.

\bibitem[{Huang et~al.(2023)Huang, Hsu, Hsu, Li, and Chen}]{huang-etal-2023-converser}
Chao-Wei Huang, Chen-Yu Hsu, Tsu-Yuan Hsu, Chen-An Li, and Yun-Nung Chen. 2023.
\newblock \href {https://doi.org/10.18653/v1/2023.sigdial-1.34} {{CONVERSER}: Few-shot conversational dense retrieval with synthetic data generation}.
\newblock In \emph{Proceedings of the 24th Annual Meeting of the Special Interest Group on Discourse and Dialogue}, pages 381--387, Prague, Czechia. Association for Computational Linguistics.

\bibitem[{Izacard and Grave(2021)}]{izacard-grave-2021-leveraging}
Gautier Izacard and Edouard Grave. 2021.
\newblock \href {https://doi.org/10.18653/v1/2021.eacl-main.74} {Leveraging passage retrieval with generative models for open domain question answering}.
\newblock In \emph{Proceedings of the 16th Conference of the European Chapter of the Association for Computational Linguistics: Main Volume}, pages 874--880, Online. Association for Computational Linguistics.

\bibitem[{Jagerman et~al.(2023)Jagerman, Zhuang, Qin, Wang, and Bendersky}]{jagerman2023query}
Rolf Jagerman, Honglei Zhuang, Zhen Qin, Xuanhui Wang, and Michael Bendersky. 2023.
\newblock Query expansion by prompting large language models.
\newblock \emph{arXiv preprint arXiv:2305.03653}.

\bibitem[{Jang et~al.(2023)Jang, Lee, Bae, Won, Lee, and Jung}]{jang2023itercqr}
Yunah Jang, Kang-il Lee, Hyunkyung Bae, Seungpil Won, Hwanhee Lee, and Kyomin Jung. 2023.
\newblock Itercqr: Iterative conversational query reformulation without human supervision.
\newblock \emph{arXiv preprint arXiv:2311.09820}.

\bibitem[{Jin et~al.(2023)Jin, Cao, Chen, Liu, and Zhao}]{jin-etal-2023-instructor}
Zhuoran Jin, Pengfei Cao, Yubo Chen, Kang Liu, and Jun Zhao. 2023.
\newblock \href {https://doi.org/10.18653/v1/2023.findings-emnlp.443} {{I}nstructo{R}: Instructing unsupervised conversational dense retrieval with large language models}.
\newblock In \emph{Findings of the Association for Computational Linguistics: EMNLP 2023}, pages 6649--6675, Singapore. Association for Computational Linguistics.

\bibitem[{Johnson et~al.(2019)Johnson, Douze, and J{\'e}gou}]{johnson2019billion}
Jeff Johnson, Matthijs Douze, and Herv{\'e} J{\'e}gou. 2019.
\newblock Billion-scale similarity search with gpus.
\newblock \emph{IEEE Transactions on Big Data}, 7(3):535--547.

\bibitem[{Joshi et~al.(2017)Joshi, Choi, Weld, and Zettlemoyer}]{joshi2017triviaqa}
Mandar Joshi, Eunsol Choi, Daniel~S Weld, and Luke Zettlemoyer. 2017.
\newblock Triviaqa: A large scale distantly supervised challenge dataset for reading comprehension.
\newblock In \emph{Proceedings of the 55th Annual Meeting of the Association for Computational Linguistics (Volume 1: Long Papers)}, pages 1601--1611.

\bibitem[{Khattab et~al.(2022)Khattab, Santhanam, Li, Hall, Liang, Potts, and Zaharia}]{khattab2022demonstrate}
Omar Khattab, Keshav Santhanam, Xiang~Lisa Li, David Hall, Percy Liang, Christopher Potts, and Matei Zaharia. 2022.
\newblock Demonstrate-search-predict: Composing retrieval and language models for knowledge-intensive {NLP}.
\newblock \emph{arXiv preprint arXiv:2212.14024}.

\bibitem[{Kim et~al.(2022)Kim, Kim, Yoo, and Kang}]{kim2022generating}
Gangwoo Kim, Sungdong Kim, Kang~Min Yoo, and Jaewoo Kang. 2022.
\newblock Generating information-seeking conversations from unlabeled documents.
\newblock In \emph{Proceedings of the 2022 Conference on Empirical Methods in Natural Language Processing}, pages 2362--2378.

\bibitem[{Kim et~al.(2023)Kim, Bae, Shin, Kang, Kwak, Yoo, and Seo}]{kim2023aligning}
Sungdong Kim, Sanghwan Bae, Jamin Shin, Soyoung Kang, Donghyun Kwak, Kang~Min Yoo, and Minjoon Seo. 2023.
\newblock Aligning large language models through synthetic feedback.
\newblock \emph{arXiv preprint arXiv:2305.13735}.

\bibitem[{Kim and Kim(2022)}]{kim2022saving}
Sungdong Kim and Gangwoo Kim. 2022.
\newblock Saving dense retriever from shortcut dependency in conversational search.
\newblock In \emph{Proceedings of the 2022 Conference on Empirical Methods in Natural Language Processing}, pages 10278--10287.

\bibitem[{Kumar and Callan(2020)}]{kumar2020making}
Vaibhav Kumar and Jamie Callan. 2020.
\newblock Making information seeking easier: An improved pipeline for conversational search.
\newblock In \emph{Findings of the Association for Computational Linguistics: EMNLP 2020}, pages 3971--3980.

\bibitem[{Kwiatkowski et~al.(2019)Kwiatkowski, Palomaki, Redfield, Collins, Parikh, Alberti, Epstein, Polosukhin, Devlin, Lee et~al.}]{kwiatkowski2019natural}
Tom Kwiatkowski, Jennimaria Palomaki, Olivia Redfield, Michael Collins, Ankur Parikh, Chris Alberti, Danielle Epstein, Illia Polosukhin, Jacob Devlin, Kenton Lee, et~al. 2019.
\newblock Natural questions: a benchmark for question answering research.
\newblock \emph{Transactions of the Association for Computational Linguistics}, 7:452--466.

\bibitem[{Lai et~al.(2024)Lai, Wu, Zhang, Sun, and Zhou}]{lai2024adacqr}
Yilong Lai, Jialong Wu, Congzhi Zhang, Haowen Sun, and Deyu Zhou. 2024.
\newblock Adacqr: Enhancing query reformulation for conversational search via sparse and dense retrieval alignment.
\newblock \emph{arXiv preprint arXiv:2407.01965}.

\bibitem[{Lin et~al.(2021{\natexlab{a}})Lin, Ma, Lin, Yang, Pradeep, and Nogueira}]{10.1145/3404835.3463238}
Jimmy Lin, Xueguang Ma, Sheng-Chieh Lin, Jheng-Hong Yang, Ronak Pradeep, and Rodrigo Nogueira. 2021{\natexlab{a}}.
\newblock \href {https://doi.org/10.1145/3404835.3463238} {Pyserini: A python toolkit for reproducible information retrieval research with sparse and dense representations}.
\newblock In \emph{Proceedings of the 44th International ACM SIGIR Conference on Research and Development in Information Retrieval}, SIGIR '21, page 2356–2362, New York, NY, USA. Association for Computing Machinery.

\bibitem[{Lin et~al.(2021{\natexlab{b}})Lin, Yang, and Lin}]{lin2021contextualized}
Sheng-Chieh Lin, Jheng-Hong Yang, and Jimmy Lin. 2021{\natexlab{b}}.
\newblock Contextualized query embeddings for conversational search.
\newblock In \emph{Proceedings of the 2021 Conference on Empirical Methods in Natural Language Processing}, pages 1004--1015.

\bibitem[{Lin et~al.(2020)Lin, Yang, Nogueira, Tsai, Wang, and Lin}]{lin2020conversational}
Sheng-Chieh Lin, Jheng-Hong Yang, Rodrigo Nogueira, Ming-Feng Tsai, Chuan-Ju Wang, and Jimmy Lin. 2020.
\newblock Conversational question reformulation via sequence-to-sequence architectures and pretrained language models.
\newblock \emph{arXiv preprint arXiv:2004.01909}.

\bibitem[{Lin et~al.(2021{\natexlab{c}})Lin, Yang, Nogueira, Tsai, Wang, and Lin}]{lin2021multistage}
Sheng-Chieh Lin, Jheng-Hong Yang, Rodrigo Nogueira, Ming-Feng Tsai, Chuan-Ju Wang, and Jimmy Lin. 2021{\natexlab{c}}.
\newblock Multi-stage conversational passage retrieval: An approach to fusing term importance estimation and neural query rewriting.
\newblock \emph{ACM Transactions on Information Systems (TOIS)}, 39(4):1--29.

\bibitem[{Loper and Bird(2002)}]{loper2002nltk}
Edward Loper and Steven Bird. 2002.
\newblock Nltk: The natural language toolkit.
\newblock \emph{arXiv preprint cs/0205028}.

\bibitem[{Ma et~al.(2023)Ma, Gong, He, Zhao, and Duan}]{ma2023query}
Xinbei Ma, Yeyun Gong, Pengcheng He, Hai Zhao, and Nan Duan. 2023.
\newblock Query rewriting for retrieval-augmented large language models.
\newblock \emph{arXiv preprint arXiv:2305.14283}.

\bibitem[{Mao et~al.(2024)Mao, Deng, Chen, Mo, Liu, Sakai, and Dou}]{mao2024chatretriever}
Kelong Mao, Chenlong Deng, Haonan Chen, Fengran Mo, Zheng Liu, Tetsuya Sakai, and Zhicheng Dou. 2024.
\newblock Chatretriever: Adapting large language models for generalized and robust conversational dense retrieval.
\newblock \emph{arXiv preprint arXiv:2404.13556}.

\bibitem[{Mao et~al.(2023{\natexlab{a}})Mao, Dou, Liu, Qian, Mo, Wu, Cheng, and Cao}]{mao2023search}
Kelong Mao, Zhicheng Dou, Bang Liu, Hongjin Qian, Fengran Mo, Xiangli Wu, Xiaohua Cheng, and Zhao Cao. 2023{\natexlab{a}}.
\newblock Search-oriented conversational query editing.
\newblock In \emph{Findings of the Association for Computational Linguistics: ACL 2023}, pages 4160--4172.

\bibitem[{Mao et~al.(2023{\natexlab{b}})Mao, Dou, Mo, Hou, Chen, and Qian}]{mao-etal-2023-large}
Kelong Mao, Zhicheng Dou, Fengran Mo, Jiewen Hou, Haonan Chen, and Hongjin Qian. 2023{\natexlab{b}}.
\newblock \href {https://doi.org/10.18653/v1/2023.findings-emnlp.86} {Large language models know your contextual search intent: A prompting framework for conversational search}.
\newblock In \emph{Findings of the Association for Computational Linguistics: EMNLP 2023}, pages 1211--1225, Singapore. Association for Computational Linguistics.

\bibitem[{Mao et~al.(2022)Mao, Dou, Qian, Mo, Cheng, and Cao}]{mao2022convtrans}
Kelong Mao, Zhicheng Dou, Hongjin Qian, Fengran Mo, Xiaohua Cheng, and Zhao Cao. 2022.
\newblock Convtrans: Transforming web search sessions for conversational dense retrieval.
\newblock In \emph{Proceedings of the 2022 Conference on Empirical Methods in Natural Language Processing}, pages 2935--2946.

\bibitem[{Mao et~al.(2023{\natexlab{c}})Mao, Qian, Mo, Dou, Liu, Cheng, and Cao}]{mao2023learning}
Kelong Mao, Hongjin Qian, Fengran Mo, Zhicheng Dou, Bang Liu, Xiaohua Cheng, and Zhao Cao. 2023{\natexlab{c}}.
\newblock Learning denoised and interpretable session representation for conversational search.
\newblock In \emph{Proceedings of the ACM Web Conference 2023}, pages 3193--3202.

\bibitem[{Mao et~al.(2021)Mao, He, Liu, Shen, Gao, Han, and Chen}]{mao2021generation}
Yuning Mao, Pengcheng He, Xiaodong Liu, Yelong Shen, Jianfeng Gao, Jiawei Han, and Weizhu Chen. 2021.
\newblock Generation-augmented retrieval for open-domain question answering.
\newblock In \emph{Proceedings of the 59th Annual Meeting of the Association for Computational Linguistics and the 11th International Joint Conference on Natural Language Processing (Volume 1: Long Papers)}, pages 4089--4100.

\bibitem[{Min et~al.(2023)Min, Krishna, Lyu, Lewis, Yih, Koh, Iyyer, Zettlemoyer, and Hajishirzi}]{min2023factscore}
Sewon Min, Kalpesh Krishna, Xinxi Lyu, Mike Lewis, Wen-tau Yih, Pang~Wei Koh, Mohit Iyyer, Luke Zettlemoyer, and Hannaneh Hajishirzi. 2023.
\newblock Factscore: Fine-grained atomic evaluation of factual precision in long form text generation.
\newblock \emph{arXiv preprint arXiv:2305.14251}.

\bibitem[{Mo et~al.(2024{\natexlab{a}})Mo, Ghaddar, Mao, Rezagholizadeh, Chen, Liu, and Nie}]{mo2024chiq}
Fengran Mo, Abbas Ghaddar, Kelong Mao, Mehdi Rezagholizadeh, Boxing Chen, Qun Liu, and Jian-Yun Nie. 2024{\natexlab{a}}.
\newblock Chiq: Contextual history enhancement for improving query rewriting in conversational search.
\newblock \emph{arXiv preprint arXiv:2406.05013}.

\bibitem[{Mo et~al.(2023)Mo, Mao, Zhu, Wu, Huang, and Nie}]{mo2023convgqr}
Fengran Mo, Kelong Mao, Yutao Zhu, Yihong Wu, Kaiyu Huang, and Jian-Yun Nie. 2023.
\newblock Convgqr: Generative query reformulation for conversational search.
\newblock \emph{arXiv preprint arXiv:2305.15645}.

\bibitem[{Mo et~al.(2024{\natexlab{b}})Mo, Qu, Mao, Zhu, Su, Huang, and Nie}]{mo2024history}
Fengran Mo, Chen Qu, Kelong Mao, Tianyu Zhu, Zhan Su, Kaiyu Huang, and Jian-Yun Nie. 2024{\natexlab{b}}.
\newblock History-aware conversational dense retrieval.
\newblock \emph{arXiv preprint arXiv:2401.16659}.

\bibitem[{Mo et~al.(2024{\natexlab{c}})Mo, Yi, Mao, Qu, Huang, and Nie}]{mo2024convsdg}
Fengran Mo, Bole Yi, Kelong Mao, Chen Qu, Kaiyu Huang, and Jian-Yun Nie. 2024{\natexlab{c}}.
\newblock Convsdg: Session data generation for conversational search.
\newblock In \emph{Companion Proceedings of the ACM on Web Conference 2024}, pages 1634--1642.

\bibitem[{Neumann et~al.(2019)Neumann, King, Beltagy, and Ammar}]{neumann2019scispacy}
Mark Neumann, Daniel King, Iz~Beltagy, and Waleed Ammar. 2019.
\newblock Scispacy: fast and robust models for biomedical natural language processing.
\newblock \emph{arXiv preprint arXiv:1902.07669}.

\bibitem[{OpenAI(2023)}]{achiam2023gpt}
OpenAI. 2023.
\newblock Gpt-4 technical report.
\newblock \emph{arXiv preprint arXiv:2303.08774}.

\bibitem[{Ouyang et~al.(2022)Ouyang, Wu, Jiang, Almeida, Wainwright, Mishkin, Zhang, Agarwal, Slama, Ray et~al.}]{ouyang2022training}
Long Ouyang, Jeffrey Wu, Xu~Jiang, Diogo Almeida, Carroll Wainwright, Pamela Mishkin, Chong Zhang, Sandhini Agarwal, Katarina Slama, Alex Ray, et~al. 2022.
\newblock Training language models to follow instructions with human feedback.
\newblock \emph{Advances in Neural Information Processing Systems}, 35:27730--27744.

\bibitem[{Qian and Dou(2022)}]{qian-dou-2022-explicit}
Hongjin Qian and Zhicheng Dou. 2022.
\newblock \href {https://doi.org/10.18653/v1/2022.emnlp-main.311} {Explicit query rewriting for conversational dense retrieval}.
\newblock In \emph{Proceedings of the 2022 Conference on Empirical Methods in Natural Language Processing}, pages 4725--4737, Abu Dhabi, United Arab Emirates. Association for Computational Linguistics.

\bibitem[{Qu et~al.(2020)Qu, Yang, Chen, Qiu, Croft, and Iyyer}]{qu2020open}
Chen Qu, Liu Yang, Cen Chen, Minghui Qiu, W~Bruce Croft, and Mohit Iyyer. 2020.
\newblock Open-retrieval conversational question answering.
\newblock In \emph{Proceedings of the 43rd International ACM SIGIR conference on research and development in Information Retrieval}, pages 539--548.

\bibitem[{Rafailov et~al.(2023)Rafailov, Sharma, Mitchell, Ermon, Manning, and Finn}]{rafailov2023direct}
Rafael Rafailov, Archit Sharma, Eric Mitchell, Stefano Ermon, Christopher~D Manning, and Chelsea Finn. 2023.
\newblock Direct preference optimization: Your language model is secretly a reward model.
\newblock \emph{arXiv preprint arXiv:2305.18290}.

\bibitem[{Raffel et~al.(2020)Raffel, Shazeer, Roberts, Lee, Narang, Matena, Zhou, Li, and Liu}]{raffel2020exploring}
Colin Raffel, Noam Shazeer, Adam Roberts, Katherine Lee, Sharan Narang, Michael Matena, Yanqi Zhou, Wei Li, and Peter~J Liu. 2020.
\newblock Exploring the limits of transfer learning with a unified text-to-text transformer.
\newblock \emph{The Journal of Machine Learning Research}, 21(1):5485--5551.

\bibitem[{Sun et~al.(2023)Sun, Shen, Zhang, Zhou, Chen, Cox, Yang, and Gan}]{sun2023salmon}
Zhiqing Sun, Yikang Shen, Hongxin Zhang, Qinhong Zhou, Zhenfang Chen, David Cox, Yiming Yang, and Chuang Gan. 2023.
\newblock Salmon: Self-alignment with principle-following reward models.
\newblock \emph{arXiv preprint arXiv:2310.05910}.

\bibitem[{Tian et~al.(2023)Tian, Mitchell, Yao, Manning, and Finn}]{tian2023fine}
Katherine Tian, Eric Mitchell, Huaxiu Yao, Christopher~D Manning, and Chelsea Finn. 2023.
\newblock Fine-tuning language models for factuality.
\newblock \emph{arXiv preprint arXiv:2311.08401}.

\bibitem[{Touvron et~al.(2023)Touvron, Martin, Stone, Albert, Almahairi, Babaei, Bashlykov, Batra, Bhargava, Bhosale et~al.}]{touvron2023llama}
Hugo Touvron, Louis Martin, Kevin Stone, Peter Albert, Amjad Almahairi, Yasmine Babaei, Nikolay Bashlykov, Soumya Batra, Prajjwal Bhargava, Shruti Bhosale, et~al. 2023.
\newblock Llama 2: Open foundation and fine-tuned chat models.
\newblock \emph{arXiv preprint arXiv:2307.09288}.

\bibitem[{Vakulenko et~al.(2021)Vakulenko, Longpre, Tu, and Anantha}]{vakulenko2021question}
Svitlana Vakulenko, Shayne Longpre, Zhucheng Tu, and Raviteja Anantha. 2021.
\newblock Question rewriting for conversational question answering.
\newblock In \emph{Proceedings of the 14th ACM international conference on web search and data mining}, pages 355--363.

\bibitem[{Van~Gysel and de~Rijke(2018)}]{VanGysel2018pytreceval}
Christophe Van~Gysel and Maarten de~Rijke. 2018.
\newblock Pytrec\_eval: An extremely fast python interface to trec\_eval.
\newblock In \emph{SIGIR}. ACM.

\bibitem[{Voskarides et~al.(2020)Voskarides, Li, Ren, Kanoulas, and de~Rijke}]{voskarides2020quretec}
Nikos Voskarides, Dan Li, Pengjie Ren, Evangelos Kanoulas, and Maarten de~Rijke. 2020.
\newblock Query resolution for conversational search with limited supervision.
\newblock In \emph{Proceedings of the 43rd International ACM SIGIR conference on research and development in Information Retrieval (SIGIR)}, pages 921--930.

\bibitem[{Wang et~al.(2023)Wang, Yang, and Wei}]{wang2023query2doc}
Liang Wang, Nan Yang, and Furu Wei. 2023.
\newblock Query2doc: Query expansion with large language models.
\newblock \emph{arXiv preprint arXiv:2303.07678}.

\bibitem[{Wei et~al.(2022)Wei, Wang, Schuurmans, Bosma, Xia, Chi, Le, Zhou et~al.}]{wei2022chain}
Jason Wei, Xuezhi Wang, Dale Schuurmans, Maarten Bosma, Fei Xia, Ed~Chi, Quoc~V Le, Denny Zhou, et~al. 2022.
\newblock Chain-of-thought prompting elicits reasoning in large language models.
\newblock \emph{Advances in Neural Information Processing Systems}, 35:24824--24837.

\bibitem[{Wu et~al.(2022)Wu, Luan, Rashkin, Reitter, Hajishirzi, Ostendorf, and Tomar}]{wu2022conqrr}
Zeqiu Wu, Yi~Luan, Hannah Rashkin, David Reitter, Hannaneh Hajishirzi, Mari Ostendorf, and Gaurav~Singh Tomar. 2022.
\newblock Conqrr: Conversational query rewriting for retrieval with reinforcement learning.
\newblock In \emph{Proceedings of the 2022 Conference on Empirical Methods in Natural Language Processing}, pages 10000--10014.

\bibitem[{Xiong et~al.(2020)Xiong, Xiong, Li, Tang, Liu, Bennett, Ahmed, and Overwijk}]{xiong2020approximate}
Lee Xiong, Chenyan Xiong, Ye~Li, Kwok-Fung Tang, Jialin Liu, Paul Bennett, Junaid Ahmed, and Arnold Overwijk. 2020.
\newblock Approximate nearest neighbor negative contrastive learning for dense text retrieval.
\newblock \emph{arXiv preprint arXiv:2007.00808}.

\bibitem[{Ye et~al.(2023)Ye, Fang, Li, and Yilmaz}]{ye2023enhancing}
Fanghua Ye, Meng Fang, Shenghui Li, and Emine Yilmaz. 2023.
\newblock Enhancing conversational search: Large language model-aided informative query rewriting.
\newblock In \emph{The 2023 Conference on Empirical Methods in Natural Language Processing}.

\bibitem[{Yu et~al.(2021)Yu, Liu, Xiong, Feng, and Liu}]{yu2021few}
Shi Yu, Zhenghao Liu, Chenyan Xiong, Tao Feng, and Zhiyuan Liu. 2021.
\newblock Few-shot conversational dense retrieval.
\newblock In \emph{Proceedings of the 44th International ACM SIGIR Conference on research and development in information retrieval}, pages 829--838.

\bibitem[{Zhang et~al.(2024)Zhang, Li, Luo, Wu, Glass, and Meng}]{zhang2024adaptive}
Tianhua Zhang, Kun Li, Hongyin Luo, Xixin Wu, James Glass, and Helen Meng. 2024.
\newblock Adaptive query rewriting: Aligning rewriters through marginal probability of conversational answers.
\newblock \emph{arXiv preprint arXiv:2406.10991}.

\bibitem[{Zheng et~al.(2023)Zheng, Chiang, Sheng, Zhuang, Wu, Zhuang, Lin, Li, Li, Xing, Zhang, Gonzalez, and Stoica}]{zheng2023judging}
Lianmin Zheng, Wei-Lin Chiang, Ying Sheng, Siyuan Zhuang, Zhanghao Wu, Yonghao Zhuang, Zi~Lin, Zhuohan Li, Dacheng Li, Eric.~P Xing, Hao Zhang, Joseph~E. Gonzalez, and Ion Stoica. 2023.
\newblock \href {http://arxiv.org/abs/2306.05685} {Judging llm-as-a-judge with mt-bench and chatbot arena}.

\end{thebibliography}

\clearpage
\appendix
\hbadness=99999  

\section{Datasets}
\begin{table}[t!]
    \small
    \centering
    \begin{threeparttable}
    \def\arraystretch{1.3}%
    \begin{tabular*}{1.03\columnwidth}{l|
    S[table-format=5.0]
    S[table-format=5.0]
    S[table-format=5.0]
    S[table-format=5.0]
    S[table-format=5.0]
    }
        \toprule
        Dataset & \text{\quad Train \quad} & \multicolumn{3}{c}{\textsc{RF Collection}} \\
        \midrule \midrule
        QReCC & & \text{QR} & \text{Plan} & \text{QE} \\
        \midrule
        \# Dialogues & 10823 & 8987 & 5519 & 8987 \\
        \# Turns & 63501 & 29596 & 8817 & 29596 \\
        \midrule\midrule
        TopiOCQA & & \text{QR} & \text{Plan} & \text{QE} \\
        \midrule
        \# Dialogues & 3509 & 3508 & 3429 & 3508 \\
        \# Turns & 45450 & 24283 & 13845 & 24283 \\
        \midrule
    \end{tabular*}
    \end{threeparttable}
    \caption{
        Statistics of \textsc{RF COllection}, QReCC, and TopiOCQA. 
    }
    \label{table:dataset_statistics}
    \vspace{-3mm}
\end{table}
\label{append:dataset}
The training dataset of QReCC comprises 10,823 conversations encompassing 63,501 turns. For evaluating queries and gathering feedback from retrieval systems, we exclude turns with no gold passage label, yielding a dataset with 8,987 conversations and 29,596 turns. 

TopiOCQA consists of 3,509 conversations with 45,450 turns. Unlike QReCC where we fully utilize the dataset, we conduct our method on a subset of TopiOCQA to manage costs associated with API requests, resulting in 3,429 conversations with 13,845 turns.
Specifically, for the QR with planning prompting method, we only apply the method to turns where the number of optimal queries generated from the QR method is less than three.

\section{\textsc{RF Collection} Details}
When constructing the collection of optimal queries $\mathcal{C}_*$, we only choose rewrites whose rank is higher than 30.
For the collection of binarized comparisons, we only consider the query with a rank higher than 50 as the preferred query.
We do not pair the queries with the same rank. 
\begin{figure}[h]
    \centering
    \includegraphics[width=\columnwidth]{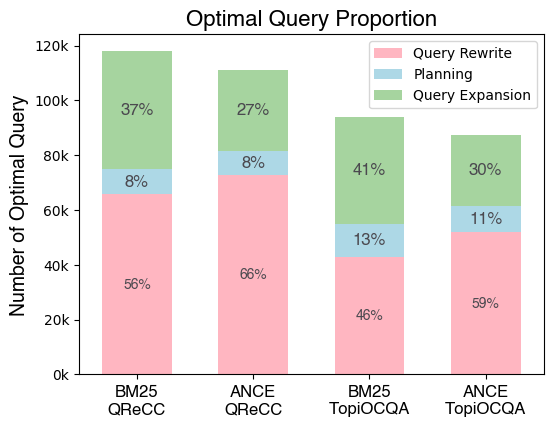}
    \caption{Proportion of optimal queries generated by each prompting method.}\vspace{-0.4cm}
    \label{fig:oq_prop}
\end{figure}
\subsection{Proportion of Question Types}
To obtain statistics in Sec. \ref{subsec:q_dist}, we use the following process.
Employing the NLTK \cite{loper2002nltk} module for query processing, part-of-speech tagging was executed, and unseen nouns and adjectives were identified through the comparison of words in the conversational history by string matching.
Queries commencing with 'what,' 'why,' 'where,' 'when,' and 'who' were categorized as Start with "\textit{Wh}" queries
Furthermore, for the categorization of queries into the query expansion style, the proportion of queries containing multiple sentences was calculated by Spacy \cite{neumann2019scispacy} library.

In Figure \ref{fig:oq_prop}, we show the proportion of query rewrite method preferences exhibited by a sparse retriever and a dense retriever on QReCC and TopiOCQA.
In the case of \textsc{RF Collection} made with feedback from BM25, It is observable that the proportion integrating the query expansion surpasses that derived from feedback by ANCE.
Moreover, within the \textsc{RF-Collection} tailored for TopiOCQA, there is an observed elevation in the number of queries generated through the query expansion and planning method in comparison to those generated from QReCC.
This tendency implies the elevated complexity inherent in TopiOCQA compared to QReCC-like topic-shifting.
The rationale behind the relatively diminished overall proportion of planning lies in its role as an auxiliary method for Query Rewrite, as previously mentioned.

\section{Experimental Details}
\begin{table*}[t!]
    \centering
    \small
    \begin{threeparttable}
    \begin{tabular*}{0.91\textwidth}{cc|cc|ccc|ccc}
        \toprule
        \multicolumn{2}{c|}{\multirow{2}{*}{\textbf{Query Reform.}}} & \multicolumn{2}{c|}{\textbf{Count}} & \multicolumn{3}{c|}{\textbf{TopiOCQA}} & \multicolumn{3}{c}{\textbf{QReCC}} \\
        \cmidrule(lr){3-4} \cmidrule(lr){5-7} \cmidrule(lr){8-10}
        & & \textbf{\#\{Calls\}} & \textbf{\#\{Queries\}} & \textbf{MRR} & \textbf{NDCG} & \textbf{R@10} & \textbf{MRR} & \textbf{NDCG} & \textbf{R@10} \\
        \midrule \midrule
        \multirow{3}{*}{\textbf{REW}} & MaxProb & \textbf{1} & \textbf{1} & 22.6 & 21.2 & 40.1 & 33.2 & 30.8 & 50.3 \\
        & Mean & 5 & 5 & 23.7 & 22.4 & 42.1 & 34.1 & 31.4 & 51.6 \\
        & SC & 5 & 5 & 23.0 & 21.9 & 40.5 & 33.5 & 31.1 & 51.6 \\
        \midrule
        \multirow{3}{*}{\textbf{RTR}} & MaxProb & 2 & 2 & 28.9 & 27.8 & 46.4 & 38.2 & 35.7 & 57.3 \\
        & Mean & 6 & 6 & 30.9 & 29.8 & 49.0 & 39.3 & 36.6 & 60.2 \\
        & SC & 6 & 6 & 29.0 & 27.9 & 47.4 & 38.1 & 35.7 & 57.6 \\
        \midrule
        \multirow{3}{*}{\textbf{RAR}} & MaxProb & \textbf{1} & 2 & 30.0 & 30.9 & 49.3 & 40.5 & 37.8 & 60.8 \\
        & Mean & 5 & 10 & 32.0 & \textbf{31.1} & 50.5 & 41.8 & 39.1 & 62.3 \\
        & SC & 5 & 10 & 31.1 & 29.9 & 50.0 & 40.6 & 38.0 & 60.8 \\
        \midrule
        RetPO (\textit{Ours}) & - & \textbf{1} & \textbf{1} & \textbf{32.2} & \textbf{31.1} & \textbf{51.6} & \textbf{44.0} & \textbf{41.1} & \textbf{66.7} \\
        \bottomrule 
    \end{tabular*}
    \end{threeparttable}
    \caption{
    Performance comparison against LLM4CS with different aggregation methods across TopiOCQA and QReCC datasets.
    We evaluate it on the ANCE retriever.
    }
    \label{table:llm4cs_updated}
    \vspace{-3mm}
\end{table*}

\paragraph{Implementation Detail}
For BM25, we set $k_1 = 0.82$, $b = 0.68$ in QReCC, and $k_1 = 0.9$, $b = 0.4$ in TopiOCQA, respectively, where $k_1$ controls the non-linear term frequency normalization and $b$ is the scale of the inverse document frequency.
We utilize GPT4-Turbo (gpt-4-1106-preview) via the OpenAI API to produce query candidates from contextualized questions. We use default hyperparameters of chat completion of API except for setting a temperature of 0.7 and maximum tokens as 1000.
For each prompting method (Question Rewriting, Planning, Query Expansion), we generate 10, 10, and 5 candidates respectively. 
We use Faiss~\cite{johnson2019billion} and Pyserini~\cite{10.1145/3404835.3463238} for efficient search across large passage indices. We retrieve top-$100$ relevant passages for each query candidate and obtain rank using pytrec\_eval~\cite{VanGysel2018pytreceval}.
Following ~\cite{kim2022saving}, the maximum token length is constrained to 128 tokens for query representations and 384 tokens for passage representations. 

We largely follow the Huggingface repository, Alignment Handbook.\footnote{\href{https://github.com/huggingface/alignment-handbook}{https://github.com/huggingface/alignment-handbook}}
We use Llama2-7b-hf as our backbone.
We use eight A100 GPUs (80GB) to train the Llama2-7b.
It is trained in one epoch for supervised fine-tuning.
We set the learning rate as 2e-5, and the batch size as 20 per GPU. The warmup ratio is set to 0.1 and we use torch data type \texttt{bfloat16}.
For the training of DPO, we set the beta as 0.1, and the maximum length as 1024.
We train our model in three epochs with a batch size of 8 per GPU.
We set the maximum input context length as 2048 and the output length as 200.

\section{Comparison with LLM4CS}
\label{appendix:llm4cs}
\begin{table*}[t!]
    \centering
    \small
    \begin{threeparttable}
    \begin{tabular*}{0.83\textwidth}{clc|ccc|ccc}
        \toprule
        & &  &  \multicolumn{3}{c}{\textbf{QReCC}} &  \multicolumn{3}{c}{\textbf{TopiOCQA}}  \\ 
        \textbf{Ret.} & \textbf{Trained on} & \textbf{Preference} & \textbf{MRR} & \textbf{R@10}  & \textbf{R@100} & \textbf{MRR} & \textbf{R@10}  & \textbf{R@100} \\
        \midrule \midrule
        \multirow{4}{*}{\rotatebox[origin=c]{90}{\textbf{BM25}}} & \multirow{2}{*}{\texttt{OQF}-QReCC} & BM25  & 50.0 & 69.5 & 89.5 & 18.1 & 31.9 & 58.7 \\
        & & ANCE & 44.4 & 66.7 & 90.0 & 17.2 & 32.0 & 59.1 \\
        \cmidrule(lr){2-9} 
        & \multirow{2}{*}{\texttt{OQF}-TopiOCQA} & BM25  & 44.7  & 66.8 & 89.3 & 28.3 & 48.3 & 73.1 \\
        & & ANCE & 40.1 & 62.2 & 86.5 & 23.1 & 41.3 & 69.4 \\
        \midrule
        \multirow{4}{*}{\rotatebox[origin=c]{90}{\textbf{ANCE}}} & \multirow{2}{*}{\texttt{OQF}-QReCC} & BM25  & 43.3 & 65.0 & 82.5 & 23.1 & 39.3 & 58.3 \\
        & & ANCE & 44.0 & 66.7 & 84.6 & 23.2 & 40.0 & 59.4 \\
        \cmidrule(lr){2-9} 
        & \multirow{2}{*}{\texttt{OQF}-TopiOCQA} & BM25  & 42.5 & 63.5 & 81.6 & 32.2 & 51.6 & 69.5 \\
        & & ANCE & 40.9 & 61.9 & 79.9 & 30.0 & 49.6 & 68.7 \\

        \bottomrule 
    \end{tabular*}
    \end{threeparttable}
    \caption{Retrieval performance when generalizing toward different setups.
    }
    \label{table:cross-pref}
    \vspace{-3mm}
\end{table*}
Table~\ref{table:llm4cs_updated} presents a detailed performance comparison of LLM4CS~\cite{mao-etal-2023-large}. We utilize \texttt{gpt-3.5-turbo-16k} as the base model, maintaining the same few-shot demonstrations as used by \citet{mao-etal-2023-large}.
The results indicate that \textsc{RetPO} consistently outperforms all combinations of prompting strategies (REW, RTR, RAR) and aggregation methods (MaxProb, Mean, SC) employed by LLM4CS, despite using only a single inference to generate high-quality queries. 
This demonstrates that our approach effectively tailors query rewrites to the retriever, achieving better performance without the need for multiple inferences.
\section{Analysis Details}
\label{append:analysis}

\begin{table}[t!]
    \small
    \centering
    \begin{threeparttable}
    \begin{tabular*}{1.01\columnwidth}{lccccc}
        \toprule
        & \multirow{2}{*}{\texttt{Orig.}} & \multicolumn{2}{c}{\textbf{RF} $\mathcal{C}_*$} & \multicolumn{2}{c}{\textbf{\textsc{RetPO}}} \\
        &  & QR & QE & Spar.  & Den. \\
        \midrule \midrule
        \# Words & 6.9 & 11.3 & 30.0 & 22.4 & 15.9 \\
        \# Unseen Words & 0.0 & 2.2 & 7.7 & 4.9 & 3.0 \\
        \% Start with `\textit{Wh}' & 62.1 & 63.5 & 0.03 & 12.8 & 28.6 \\
        \% Multiple Sents. & 0.08 & 0.2 & 99.4 & 59.8 & 27.7 \\

        \midrule\
    \end{tabular*}
    \end{threeparttable}
    \caption{
    Statistics for question distributions from \textsc{RF Collection} and \textsc{RetPO}.
    We compare the number of words and the structure of questions. 
    }
    \label{table:q_dist}
    \vspace{-5mm}
\end{table}

\begin{figure} [t!]
    \centering
    \includegraphics[width=\columnwidth]{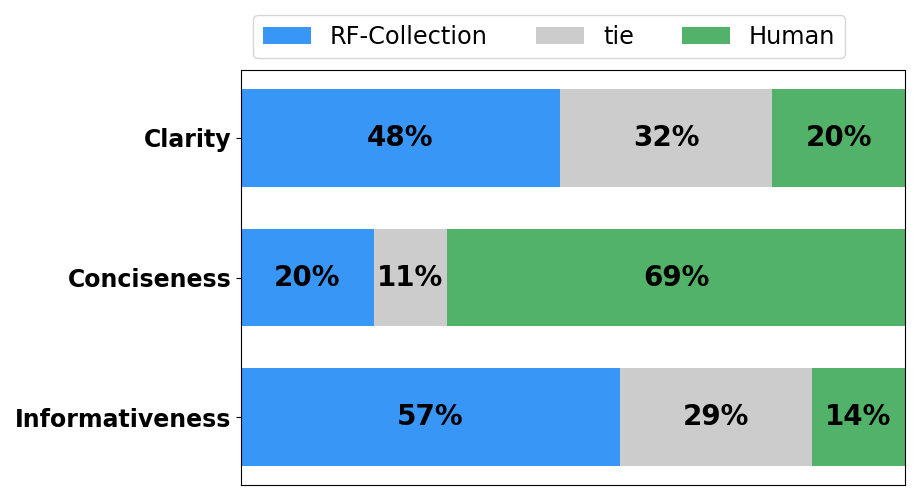}
    \caption{Pairwise evaluation with GPT-4. Rewrites from \textsc{RF Collection} are compared with the human rewrites.}\vspace{-0.4cm}
    \label{fig:gpt4_eval_rf}
\end{figure}
\subsection{Comparison of Question Distributions}
\label{subsec:q_dist}

Table \ref{table:q_dist} presents a statistical analysis of query distributions of optimal queries from \textsc{RF Collection} $\mathcal{C}_*$ and predicted rewrites from RetPO methods. 
It shows the number of words, frequency of unseen words from the held-out conversation, questions starting with `\textit{Wh}-' words, and those composed of multiple sentences.\footnote{See Appendix \ref{append:analysis} for details about the measurements.}
\textsc{RF Collection} and \textsc{RetPO} tend to create longer queries, often extending to 2-5 times the length of the original one, which includes a number of words unseen within the utterances so far.
The query expansion (QE) notably alters the question structure, frequently constructing them as multi-sentence entities (high \% of Multiple Sents.). 
This method tends to prepend a pseudo-answer to the question (low \% of Starting with '\textit{Wh-}').
\textsc{RetPO}, in contrast, strikes a balance between QR and QE, achieving a midpoint depending on the retriever type.

\subsection{GPT-4 Evaluation Details}
\label{appendix:GPT-4_eval_details}

Prompts used in GPT-4 evaluation are shown in Table~\ref{table:gpt4_eval_clarity}, \ref{table:gpt4_eval_conciseness}, and \ref{table:gpt4_eval_informativeness}. Considering the position bias in GPT-4 evaluation~\cite{zheng2023judging}, we assess the same instance twice, reversing the order of the two rewritten questions. Also, we regard the comparison as a `Tie' if the two evaluation results conflict with each other.

\subsection{Evaluating \textsc{RF Collection}}
\label{appendix:eval_dataset}
In Table \ref{fig:gpt4_eval_rf}, we present a comprehensive comparison of various query reformulation strategies. 
We assess the performance of rewrites generated from our \textsc{RF Collection} against baselines including oracle setups. 
We report the best retrieval performances of each set.
All of our prompting methods significantly outperform \texttt{Human Rewrite} with a huge gap in most metrics.
Query expansion shows the best performance among the prompting methods, showing its efficacy in adding keywords.
The combined set \texttt{Union} of all strategies yields the best results, indicating these methods are mutually beneficial.

\begin{table}[t!]
    \small
    \centering
    \begin{threeparttable}
    \begin{tabular*}{1.0\columnwidth}{lcccc}
        \toprule
        \textbf{Query Reform.} & \textbf{\#(Q)} &  \textbf{MRR} & \textbf{R@10}  & \textbf{R@100} \\
        \midrule \midrule
        \multicolumn{5}{c}{\textit{Sparse (BM25)}} \\
        \midrule
        \texttt{Original} & 1 & 6.5 & 11.1 & 21.5 \\
        \texttt{Concat} ($H_{<t}, q_t$) & 1 & 47.0 & 65.1 & 82.8 \\
        \texttt{Human Rewrite} & 1 & 40.0 & 62.7 & 98.5 \\
        \quad \texttt{+ Gold Answer} & 1 & 92.4 & 97.2 & 99.7 \\
        \midrule
        \multicolumn{4}{l}{\textsc{RF Collection}} \\
        \quad Query Rewriting& 10 & 64.5 & 81.1 & 94.5 \\
        \quad \quad w/ Planning & 10 & 68.2 & 83.6 & 95.2 \\
        \quad Query Expansion & 5 & 75.0 & 91.3 & 99.1 \\
        \quad \texttt{Union} & 25 & 85.1 & 93.7 & 98.6 \\

        \midrule\
    \end{tabular*}
    \end{threeparttable}
    \caption{
    Comparison of effectiveness with BM25 over different query reformulation strategies. 
    We evaluate the performance of our generated rewrites from \textsc{RF Collection} against simple baselines and oracle setups.
    }
    \label{table:oqf_bm25}
    \vspace{-3mm}
\end{table}
\section{Case Study}
In Table~\ref{table:case_study_qrecc_bm25_success}, we demonstrate the effectiveness of \textsc{RetPO} in enhancing retrieval performance by providing additional specific information. While the information generated by \textsc{RetPO} does not seemingly overlap with the actual answer, they nevertheless contribute by offering supplementary cues that guide the retriever toward the most pertinent passages.
\subsection{Over-specification Issue}
\label{appendix:over_spec}
In Table~\ref{table:case_study_qrecc_bm25_fail}, we present a failure case where \textsc{RetPO} fails to accurately align with the original search intent, resulting in a misjudgment during retrieval. The deviation from the original question scope is highlighted, indicating an over-specification in the output query. This over-specification leads to a mismatch with the intended search query, thereby hindering successful retrieval.

\section{Prompts}
Table~\ref{table:prompt_qr},~\ref{table:prompt_plan},~\ref{table:prompt_qe} illustrate examples of our prompting methods: question rewriting (QR), QR with planning, and query expansion. Following ~\cite{khattab2022demonstrate}, each prompt comprises four components: an instruction, a format specification, a few-shot example, and a test instance. 
In question rewriting, we instruct LLM to generate a series of decontextualized questions adhering to the predefined criteria proposed by ~\cite{ye2023enhancing}. 
In QR with planning, the LLM is guided to elicit relevant information that might help reformulate a question, before generating each rewritten question. 
In query expansion, LLM produces a set of pseudo-answer candidates expected to align closely with the potential response of the question. 
We use a one-shot example for each prompting method to demonstrate the desired action and output.

\begin{figure}[t!]
    \centering
    \includegraphics[width=\columnwidth]{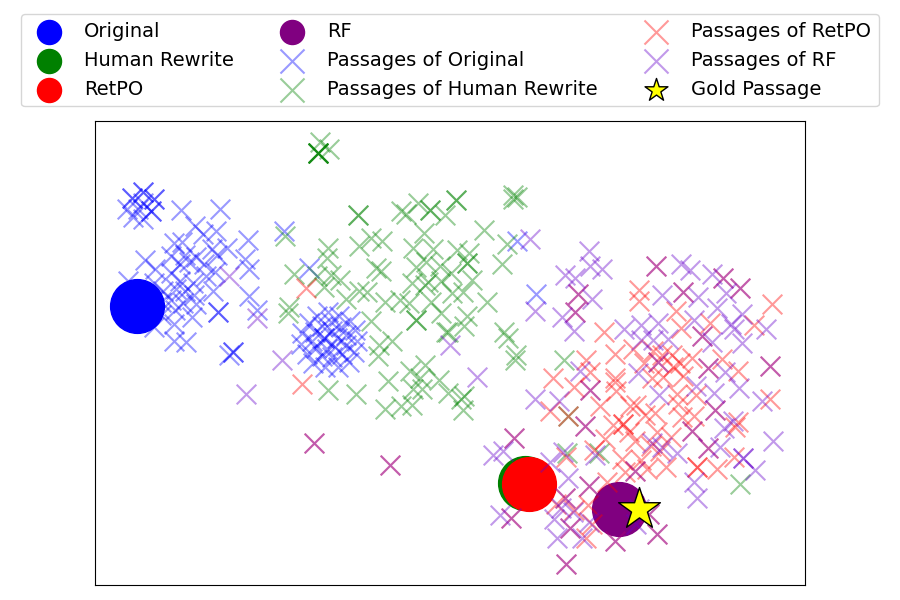}
    \caption{T-SNE visualization of ANCE embeddings from \textsc{RetPO} and \textsc{RF Collection}. 
    Queries and passages from the same method are colored identically. }\vspace{-0.4cm}
    \label{fig:tsne}
\end{figure}
\clearpage

\begin{table*}[t!]
    \centering
    \resizebox{\textwidth}{!}{
    \def\arraystretch{0.9}
    \begin{tabularx}{\textwidth}{X}
        \toprule
            \textbf{[Instruction]}\\
            Please act as an impartial judge and evaluate the quality of the query-rewriting system displayed below.
            The system tries to rewrite the conversational input to a stand-alone question, eliminating dependency on the conversational context.\\\\
            Your job is to compare the \textbf{clarity} of the two rewritten stand-alone questions.\\
            That is, You should check which question is \textbf{less open to multiple interpretations and has a more clear intention.}\\
            Please choose either 'A' or 'B'. If the two questions show the same clarity, answer it by 'Tie'. For example, Judge: (A|B|Tie)\\\\            
            \text{[Conversation]}\\
            \{\textit{conversation}\}\\\\
            
            \text{[The Start of stand-alone question A]} \\
            \{\textit{query}\_1\}\\
            \text{[The End of stand-alone question A]} \\\\
            
            \text{[The Start of stand-alone question B]} \\
            \{\textit{query\_2}\}\\
            \text{[The End of stand-alone question B]} \\\\
            
            Judge: \\
        \bottomrule
    \end{tabularx}}
    \caption{
    GPT4 prompt for evaluating clarity
    }
    \label{table:gpt4_eval_clarity}
    \vspace{-.4cm}
\end{table*}
\begin{table*}[t!]
    \centering
    \resizebox{\textwidth}{!}{
    \def\arraystretch{0.9}
    \begin{tabularx}{\textwidth}{X}
        \toprule
            \textbf{[Instruction]}\\
            Please act as an impartial judge and evaluate the quality of the query-rewriting system displayed below.
            The system tries to rewrite the conversational input to a stand-alone question, eliminating dependency on the conversational context.\\\\
            Your job is to compare the \textbf{conciseness} of the two rewritten stand-alone questions.\\
            That is, You should check which question is \textbf{more brief and directly states the search intent without additional elaboration.}\\
            Please choose either 'A' or 'B'. If the two questions show the same conciseness, answer it by 'Tie'. For example, Judge: (A|B|Tie)\\\\        
            \text{[Conversation]}\\
            \{\textit{conversation}\}\\\\
            
            \text{[The Start of stand-alone question A]} \\
            \{\textit{query}\_1\}\\
            \text{[The End of stand-alone question A]} \\\\
            
            \text{[The Start of stand-alone question B]} \\
            \{\textit{query\_2}\}\\
            \text{[The End of stand-alone question B]} \\\\
            
            Judge:\\
        \bottomrule 
    \end{tabularx}}
    \caption{
    GPT4 prompt for evaluating conciseness
    }
    \label{table:gpt4_eval_conciseness}
    \vspace{-.4cm}
\end{table*}
\begin{table*}[t!]
    \centering
    \resizebox{\textwidth}{!}{
    \def\arraystretch{0.9}
    \begin{tabularx}{\textwidth}{X}
        \toprule
            \textbf{[Instruction]}\\
            Please act as an impartial judge and evaluate the quality of the query-rewriting system displayed below.
            The system tries to rewrite the conversational input to a stand-alone question, eliminating dependency on the conversational context.\\\\
            Your job is to compare the \textbf{informativeness} of the two rewritten stand-alone questions.\\
            That is, You should check \textbf{which question provides more useful and relevant information.}\\
            Please choose either 'A' or 'B'. If the two questions show the same informativeness, answer it by 'Tie'. For example, Judge: (A|B|Tie)\\\\        
            \text{[Conversation]}\\
            \{\textit{conversation}\}\\\\
            
            \text{[The Start of stand-alone question A]} \\
            \{\textit{query}\_1\}\\
            \text{[The End of stand-alone question A]} \\\\
            
            \text{[The Start of stand-alone question B]} \\
            \{\textit{query\_2}\}\\
            \text{[The End of stand-alone question B]} \\\\
            
            Judge:\\
        \bottomrule 
    \end{tabularx}}
    \caption{
    GPT4 prompt for evaluating informativeness
    }
    \label{table:gpt4_eval_informativeness}
    \vspace{-.4cm}
\end{table*}
\begin{table*}[t!]
    \centering
    \resizebox{\textwidth}{!}{
    \def\arraystretch{0.9}
    \begin{tabularx}{\textwidth}{X}
        \toprule
            \textbf{Conversation}:\\
            Q1: where are we now video who is the girl
            A1: The Where Are We Now? music video, directed by Tony Oursler, shows Bowie and an unnamed female companion. The woman was later confirmed as artist Jacqueline Humphries, Oursler's wife. \\
            Q2: why was the wife in the music video
            A2: David Bowie and Oursler reportedly wanted someone who looked like Bowie's PA, Corinne Coco Schwab, as she did in the 1970s in Berlin. \\
            Q3: where is the setting of the music video
            A3: The video is set in what could be an artists' studio in Berlin, where Bowie lived from 1976, showing moving black-and-white footage of the city from the 1970s.\\
            Q4: any references in the music video
            A4: The footage on the screen and references in the lyrics include the Berlin Wall and mention of the Bosebrucke, the first border crossing that opened when the Wall fell.\\\\
            \textbf{Original}: any other references (\textbf{rank: Not Found}) \\\\
            \textbf{Human Rewrite}: any other references in the where are we now music video besides the berlin wall and bosebrucke (\textbf{rank: Not Found})\\\\
            \textbf{\textsc{RetPO}}: The video may also include references to Bowie's time in Berlin during the 1970s, such as the black-and-white footage of the city and the mention of the Bosebrucke.\textbackslash{n}Any other references in the Where Are We Now music video besides the Berlin Wall and mention of the Bosebrucke (\textbf{rank: 1})\\\\
            \textbf{Answer}: Other reference points in the video and song are the Brandenburg Gate\; the Dschungel nightclub; the Fernsehturm, or television tower\; KaDeWe, the department store\; and graffiti from Kunsthaus Tacheles. \\
        \bottomrule 
    \end{tabularx}}
    \caption{
    Success case on QReCC (2396\_5) when using BM25.
    }
    \label{table:case_study_qrecc_bm25_success}
    \vspace{-.4cm}
\end{table*}

\begin{table*}[t!]
    \centering
    \resizebox{\textwidth}{!}{
    \def\arraystretch{0.9}
    \begin{tabularx}{\textwidth}{X}
        \toprule
            \textbf{Conversation}:\\
            Q1: what is mechanical animals?
            A1: Mechanical Animals is the third studio album by American rock band Marilyn Manson. \\
            Q2: what kind of music did this album have?
            A2: A rock opera and concept album, Mechanical Animals is the second installment in a trilogy \\
            Q3: who producedthe album?
            A3: Marilyn Manson subsequently employed Michael Beinhorn as principal producer, co-producing the record with Marilyn Manson. Sean Beavan was also brought in to supply additional production work. \\
            Q4: was this album a hit on the charts?
            A4: Mechanical Animals debuted at number one on the Billboard 200 with first-week sales of 223,000 units, becoming Marilyn Manson's first number-one album on the chart. \\
            Q5: did it win any awards?
            A5: Mechanical Animals was certified platinum by the Recording Industry Association of America on February 25, 1999, \\
            Q6: did they go on tour?
            A6: The Beautiful Monsters Tour was a North American concert tour co-headlined by American rock bands Hole and Marilyn Manson. \\\\
            \textbf{Original}: who was hole? (\textbf{rank: Not Found})\\\\
            \textbf{Human Rewrite}: Who was American rock band Hole? (\textbf{rank: 3})\\\\
            \textbf{\textsc{RetPO}}: Who were the \textcolor{red}{members of the band Hole} that co-headlined the Beautiful Monsters Tour with Marilyn Manson? (\textbf{rank: Not Found})\\\\
            \textbf{Answer}: Hole was an American alternative rock band formed in Los Angeles, California in 1989.\\
        \bottomrule 
    \end{tabularx}}
    \caption{
    Failure case in QReCC (1321\_7) when using BM25. The \textcolor{red}{red text} indicates the deviation from the original question scope. 
    The resulting query from \textsc{RetPO} over-specifies irrelevant details, asking about members of the band Hole, rather than the band as a whole. It leads to misalignment with the original search intent. 
    }
    \label{table:case_study_qrecc_bm25_fail}
    \vspace{-.4cm}
\end{table*}
\begin{table*}[t!]
    \centering
    \resizebox{\textwidth}{!}{
    \def\arraystretch{0.9}
    \begin{tabularx}{\textwidth}{X}
        \toprule
            Given a question and its context, decontextualize the question by addressing coreference and omission issues. The resulting question should retain its original meaning and be as informative as possible, and should not duplicate any previously asked questions in the context. Please give me a list of 10 candidates
            for the rewrite. Here are some examples. \\\\
            - - -\\\\
            Follow the following format.\\\\
            \textbf{Conversation}:\\
            \$\{conversational context for the question\}\\\\
            \textbf{Question}:
            \$\{follow-up question to be rewritten\}\\\\
            \textbf{Rewrite}:
            \$\{list of 10 rewritten question candidates, each on a new line.\}\\
            Rewrite i:
            \$\{(i)-th rewritten question that address coreference and omission issues\} \\\\
            - - -\\\\
            \textbf{Conversation}:
            \\ Q1: How did religion effect their society? A1: Religion held ancient Hawaiian society together, affecting habits, lifestyles, work methods, social policy and law. The legal system was based on religious kapu, or taboos.
            \\ Q2: What is Kapu? A2: Kapu is the ancient Hawaiian code of conduct of laws and regulations.
            \\ ...
            \\ Q4: What are the beginnings of the kapu system like? A4: The rigidity of the kapu system might have come from a second wave of migrations in 1000–1300 from which different religions and systems were shared\\\\
            \textbf{Question}: How did this wave effect society or the system?\\\\
            \textbf{Rewrite}:\\
            \texttt{\textcolor{teal}{Rewrite 1: How did the second wave of migrations between 1000–1300 impact ancient Hawaiian society or the kapu system?}}\\
            \texttt{\textcolor{teal}{Rewrite 2: In what ways were the social structure or kapu system of ancient Hawaii influenced by migrations from 1000 to 1300?}}\\
            \texttt{\textcolor{teal}{Rewrite 3: ...}}\\
        \bottomrule 
    \end{tabularx}}
    \caption{
    Prompt for the question rewriting method
    }
    \label{table:prompt_qr}
    \vspace{-.4cm}
\end{table*}
\begin{table*}[t!]
    \centering
    \resizebox{\textwidth}{!}{
    \def\arraystretch{0.9}
    \begin{tabularx}{\textwidth}{X}
        \toprule
            I am working on finding information to rewrite the question. Given a question and its context, Please provide 10 information-Rewrite pairs, where each pair consists of information that might be needed to answer the question and a rewritten question. the rewritten question is a decontextualized version of the question by addressing coreference and omission issues with respect to each information. the resulting question should retain its original search intent. Here are some examples.\\\\
            - - -\\\\
            Follow the following format.\\\\
            \textbf{Conversation}:\\
            \$\{conversational context for the question\}\\\\
            \textbf{Question}:
            \$\{follow-up question to be rewritten\}\\\\
            \textbf{Information-Rewrite:}
            \$\{list of 10 Information-Rewrite pairs, each on a new line\}\\
            Info i:
            \$\{(i)-th information that is needed to answer the question. it should not be too specific\}\\
            Rewrite i:
            \$\{(i)-th rewritten question that address coreference and omission issues with respect to (i)-th information.\}\\\\
            - - -\\\\
            \textbf{Conversation}:\\
            Q1: How did religion effect their society? A1: Religion held ancient Hawaiian society together, affecting habits, lifestyles, work methods, social policy and law. The legal system was based on religious kapu, or taboos.\\
            Q2: What is Kapu? A2: Kapu is the ancient Hawaiian code of conduct of laws and regulations.
            \\ ...
            \\ Q4: What are the beginnings of the kapu system like? A4: The rigidity of the kapu system might have come from a second wave of migrations in 1000–1300 from which different religions and systems were shared\\\\
            \textbf{Question}: How did this wave effect society or the system?\\\\
            \textbf{Information-Rewrite}:\\
            \texttt{\textcolor{teal}{Info 1: Migration Impact - Information about how the second wave of migrations influenced the existing societal structures or introduced changes in ancient Hawaiian society.}}\\
            \textbf{\texttt{\textcolor{teal}{Rewrite 1: How did the second wave of migrations around 1000-1300 AD affect ancient Hawaiian society and its structures?}}}\\\\
            \texttt{\textcolor{teal}{Info 2: Changes to Kapu System - Details regarding any modifications or introductions to the kapu system as a result of the second wave of migrations.}}\\
            \textbf{\texttt{\textcolor{teal}{Rewrite 2: What changes were made to the ancient Hawaiian kapu system due to the second wave of migrations?}}}\\\\
            \texttt{\textcolor{teal}{Info 3: ...}} \\
        \bottomrule 
    \end{tabularx}}
    \caption{
    Prompt for the planning method.
    }
    \label{table:prompt_plan}
    \vspace{-.4cm}
\end{table*}
\begin{table*}[t!]
    \centering
    \resizebox{\textwidth}{!}{
    \def\arraystretch{0.9}
    \begin{tabularx}{\textwidth}{X}
        \toprule
            Please give me a list of 5 answer candidates based on the given conversation context and question. Here are some examples. \\\\
            - - -\\\\
            Follow the following format.\\\\
            \textbf{Conversation}:\\
            \$\{conversational context for the question\}\\\\
            \textbf{Question}:
            \$\{follow-up question to be rewritten\}\\\\
            \textbf{Answer}:
            \$\{list of 5 answer candidates, each on a new line.\}\\
            Answer i:
            \$\{(i)-th answer for the current question\} \\\\
            - - -\\\\
            \textbf{Conversation}:
            \\ Q1: How did religion effect their society? A1: Religion held ancient Hawaiian society together, affecting habits, lifestyles, work methods, social policy and law. The legal system was based on religious kapu, or taboos.
            \\ Q2: What is Kapu? A2: Kapu is the ancient Hawaiian code of conduct of laws and regulations.
            \\ ...
            \\ Q4: What are the beginnings of the kapu system like? A4: The rigidity of the kapu system might have come from a second wave of migrations in 1000–1300 from which different religions and systems were shared\\\\
            \textbf{Question}: How did this wave effect society or the system?\\\\
            \textbf{Answer}:\\
            \texttt{\textcolor{teal}{Answer 1: The second wave of migrations brought new religious beliefs and practices, which likely intensified the existing kapu system and introduced additional taboos.}}\\
            \texttt{\textcolor{teal}{Answer 2: The influx of migrants during this period could have led to the formalization and expansion of the kapu system, as new ideas were integrated and enforced.}}\\
            \texttt{\textcolor{teal}{Answer 3: ...}}\\
        \bottomrule 
    \end{tabularx}}
    \caption{
    Prompt for the query expansion method. We concatenate the pseudo-answers with a self-contained query.
    }
    \label{table:prompt_qe}
    \vspace{-.4cm}
\end{table*}

\end{document}